\documentclass[aps,prd,twocolumn,superscriptaddress,showpacs,letterpaper]{revtex4}

\usepackage{graphicx}
\usepackage{amsmath, amssymb}
\usepackage{dcolumn}
\usepackage{bm}
\usepackage{xspace}
\usepackage{multirow}
\usepackage{booktabs}
\usepackage{longtable}
\usepackage{textcomp}
\usepackage{subfigure}

\usepackage{definitions}

\begin{document}

\title{Measurements of the properties of $\Lambda_c(2595)$,
  $\Lambda_c(2625)$, $\Sigma_c(2455)$, and $\Sigma_c(2520)$ baryons}

\affiliation{Institute of Physics, Academia Sinica, Taipei, Taiwan 11529, Republic of China} 
\affiliation{Argonne National Laboratory, Argonne, Illinois 60439, USA} 
\affiliation{University of Athens, 157 71 Athens, Greece} 
\affiliation{Institut de Fisica d'Altes Energies, ICREA, Universitat Autonoma de Barcelona, E-08193, Bellaterra (Barcelona), Spain} 
\affiliation{Baylor University, Waco, Texas 76798, USA} 
\affiliation{Istituto Nazionale di Fisica Nucleare Bologna, $^{aa}$University of Bologna, I-40127 Bologna, Italy} 
\affiliation{University of California, Davis, Davis, California 95616, USA} 
\affiliation{University of California, Los Angeles, Los Angeles, California 90024, USA} 
\affiliation{Instituto de Fisica de Cantabria, CSIC-University of Cantabria, 39005 Santander, Spain} 
\affiliation{Carnegie Mellon University, Pittsburgh, Pennsylvania 15213, USA} 
\affiliation{Enrico Fermi Institute, University of Chicago, Chicago, Illinois 60637, USA}
\affiliation{Comenius University, 842 48 Bratislava, Slovakia; Institute of Experimental Physics, 040 01 Kosice, Slovakia} 
\affiliation{Joint Institute for Nuclear Research, RU-141980 Dubna, Russia} 
\affiliation{Duke University, Durham, North Carolina 27708, USA} 
\affiliation{Fermi National Accelerator Laboratory, Batavia, Illinois 60510, USA} 
\affiliation{University of Florida, Gainesville, Florida 32611, USA} 
\affiliation{Laboratori Nazionali di Frascati, Istituto Nazionale di Fisica Nucleare, I-00044 Frascati, Italy} 
\affiliation{University of Geneva, CH-1211 Geneva 4, Switzerland} 
\affiliation{Glasgow University, Glasgow G12 8QQ, United Kingdom} 
\affiliation{Harvard University, Cambridge, Massachusetts 02138, USA} 
\affiliation{Division of High Energy Physics, Department of Physics, University of Helsinki and Helsinki Institute of Physics, FIN-00014, Helsinki, Finland} 
\affiliation{University of Illinois, Urbana, Illinois 61801, USA} 
\affiliation{The Johns Hopkins University, Baltimore, Maryland 21218, USA} 
\affiliation{Institut f\"{u}r Experimentelle Kernphysik, Karlsruhe Institute of Technology, D-76131 Karlsruhe, Germany} 
\affiliation{Center for High Energy Physics: Kyungpook National University, Daegu 702-701, Korea; Seoul National University, Seoul 151-742, Korea; Sungkyunkwan University, Suwon 440-746, Korea; Korea Institute of Science and Technology Information, Daejeon 305-806, Korea; Chonnam National University, Gwangju 500-757, Korea; Chonbuk National University, Jeonju 561-756, Korea} 
\affiliation{Ernest Orlando Lawrence Berkeley National Laboratory, Berkeley, California 94720, USA} 
\affiliation{University of Liverpool, Liverpool L69 7ZE, United Kingdom} 
\affiliation{University College London, London WC1E 6BT, United Kingdom} 
\affiliation{Centro de Investigaciones Energeticas Medioambientales y Tecnologicas, E-28040 Madrid, Spain} 
\affiliation{Massachusetts Institute of Technology, Cambridge, Massachusetts 02139, USA} 
\affiliation{Institute of Particle Physics: McGill University, Montr\'{e}al, Qu\'{e}bec, Canada H3A~2T8; Simon Fraser University, Burnaby, British Columbia, Canada V5A~1S6; University of Toronto, Toronto, Ontario, Canada M5S~1A7; and TRIUMF, Vancouver, British Columbia, Canada V6T~2A3} 
\affiliation{University of Michigan, Ann Arbor, Michigan 48109, USA} 
\affiliation{Michigan State University, East Lansing, Michigan 48824, USA}
\affiliation{Institution for Theoretical and Experimental Physics, ITEP, Moscow 117259, Russia}
\affiliation{University of New Mexico, Albuquerque, New Mexico 87131, USA} 
\affiliation{Northwestern University, Evanston, Illinois 60208, USA} 
\affiliation{The Ohio State University, Columbus, Ohio 43210, USA} 
\affiliation{Okayama University, Okayama 700-8530, Japan} 
\affiliation{Osaka City University, Osaka 588, Japan} 
\affiliation{University of Oxford, Oxford OX1 3RH, United Kingdom} 
\affiliation{Istituto Nazionale di Fisica Nucleare, Sezione di Padova-Trento, $^{bb}$University of Padova, I-35131 Padova, Italy} 
\affiliation{LPNHE, Universite Pierre et Marie Curie/IN2P3-CNRS, UMR7585, Paris, F-75252 France} 
\affiliation{University of Pennsylvania, Philadelphia, Pennsylvania 19104, USA}
\affiliation{Istituto Nazionale di Fisica Nucleare Pisa, $^{cc}$University of Pisa, $^{dd}$University of Siena and $^{ee}$Scuola Normale Superiore, I-56127 Pisa, Italy} 
\affiliation{University of Pittsburgh, Pittsburgh, Pennsylvania 15260, USA} 
\affiliation{Purdue University, West Lafayette, Indiana 47907, USA} 
\affiliation{University of Rochester, Rochester, New York 14627, USA} 
\affiliation{The Rockefeller University, New York, New York 10065, USA} 
\affiliation{Istituto Nazionale di Fisica Nucleare, Sezione di Roma 1, $^{ff}$Sapienza Universit\`{a} di Roma, I-00185 Roma, Italy} 

\affiliation{Rutgers University, Piscataway, New Jersey 08855, USA} 
\affiliation{Texas A\&M University, College Station, Texas 77843, USA} 
\affiliation{Istituto Nazionale di Fisica Nucleare Trieste/Udine, I-34100 Trieste, $^{gg}$University of Udine, I-33100 Udine, Italy} 
\affiliation{University of Tsukuba, Tsukuba, Ibaraki 305, Japan} 
\affiliation{Tufts University, Medford, Massachusetts 02155, USA} 
\affiliation{University of Virginia, Charlottesville, Virginia 22906, USA}
\affiliation{Waseda University, Tokyo 169, Japan} 
\affiliation{Wayne State University, Detroit, Michigan 48201, USA} 
\affiliation{University of Wisconsin, Madison, Wisconsin 53706, USA} 
\affiliation{Yale University, New Haven, Connecticut 06520, USA} 
\author{T.~Aaltonen}
\affiliation{Division of High Energy Physics, Department of Physics, University of Helsinki and Helsinki Institute of Physics, FIN-00014, Helsinki, Finland}
\author{B.~\'{A}lvarez~Gonz\'{a}lez$^w$}
\affiliation{Instituto de Fisica de Cantabria, CSIC-University of Cantabria, 39005 Santander, Spain}
\author{S.~Amerio}
\affiliation{Istituto Nazionale di Fisica Nucleare, Sezione di Padova-Trento, $^{bb}$University of Padova, I-35131 Padova, Italy} 

\author{D.~Amidei}
\affiliation{University of Michigan, Ann Arbor, Michigan 48109, USA}
\author{A.~Anastassov}
\affiliation{Northwestern University, Evanston, Illinois 60208, USA}
\author{A.~Annovi}
\affiliation{Laboratori Nazionali di Frascati, Istituto Nazionale di Fisica Nucleare, I-00044 Frascati, Italy}
\author{J.~Antos}
\affiliation{Comenius University, 842 48 Bratislava, Slovakia; Institute of Experimental Physics, 040 01 Kosice, Slovakia}
\author{G.~Apollinari}
\affiliation{Fermi National Accelerator Laboratory, Batavia, Illinois 60510, USA}
\author{J.A.~Appel}
\affiliation{Fermi National Accelerator Laboratory, Batavia, Illinois 60510, USA}
\author{A.~Apresyan}
\affiliation{Purdue University, West Lafayette, Indiana 47907, USA}
% ...
\collaboration{CDF Collaboration\footnote{With visitors from $^a$University of MA Amherst,
Amherst, MA 01003, USA,
$^b$Istituto Nazionale di Fisica Nucleare, Sezione di Cagliari, 09042 Monserrato (Cagliari), Italy,
$^c$University of CA Irvine, Irvine, CA  92697, USA,
$^d$University of CA Santa Barbara, Santa Barbara, CA 93106, USA,
$^e$University of CA Santa Cruz, Santa Cruz, CA  95064, USA,
$^f$CERN,CH-1211 Geneva, Switzerland,
$^g$Cornell University, Ithaca, NY  14853, USA, 
$^h$University of Cyprus, Nicosia CY-1678, Cyprus, 
$^i$University College Dublin, Dublin 4, Ireland,
$^j$University of Fukui, Fukui City, Fukui Prefecture, Japan 910-0017,
$^k$Universidad Iberoamericana, Mexico D.F., Mexico,
$^l$Iowa State University, Ames, IA  50011, USA,
$^m$University of Iowa, Iowa City, IA  52242, USA,
$^n$Kinki University, Higashi-Osaka City, Japan 577-8502,
$^o$Kansas State University, Manhattan, KS 66506, USA,
$^p$University of Manchester, Manchester M13 9PL, United Kingdom,
$^q$Queen Mary, University of London, London, E1 4NS, United Kingdom,
$^r$University of Melbourne, Victoria 3010, Australia,
$^s$Muons, Inc., Batavia, IL 60510, USA,
$^t$Nagasaki Institute of Applied Science, Nagasaki, Japan, 
$^u$National Research Nuclear University, Moscow, Russia,
$^v$University of Notre Dame, Notre Dame, IN 46556, USA,
$^w$Universidad de Oviedo, E-33007 Oviedo, Spain, 
$^x$Texas Tech University, Lubbock, TX  79609, USA,
$^y$Universidad Tecnica Federico Santa Maria, 110v Valparaiso, Chile,
$^z$Yarmouk University, Irbid 211-63, Jordan,
$^{ii}$University of Warwick, Coventry CV4 7AL, United Kingdom,
$^{hh}$On leave from J.~Stefan Institute, Ljubljana, Slovenia, 
}}
\noaffiliation

\noaffiliation

\pacs{14.20.Lq, 14.20.Gk}

\begin{abstract}
We report measurements of the resonance properties of
$\Lambda_c(2595)^+$ and $\Lambda_c(2625)^+$ baryons in their decays to
$\Lambda_c^+\pi^+\pi^-$ as well as $\Sigma_c(2455)^{++,0}$ and
$\Sigma_c(2520)^{++,0}$ baryons in their decays to
$\Lambda_c^+\pi^\pm$ final states. These measurements are performed
using data corresponding to 5.2\,\invfb of integrated luminosity from
$p\bar{p}$ collisions at $\sqrt{s}$ = 1.96\,TeV, collected with the
CDF II detector at the Fermilab Tevatron. Exploiting the largest
available charmed baryon sample, we measure masses and decay widths
with uncertainties comparable to the world averages for $\Sigma_c$
states, and significantly smaller uncertainties than the world
averages for excited $\Lambda_c^+$ states.
\end{abstract}

\maketitle

\section{Introduction}
\label{sec:Intro}

Hadrons containing a $b$ or $c$ quark are referred to as heavy-quark
hadrons and provide an interesting laboratory for studying and testing
quantum chromodynamics (QCD), the theory of strong
interactions~\cite{PhysRevLett.30.1343,PhysRevLett.30.1346}. Because
the strong coupling constant $\alpha_s$ is large for interactions
involving small momentum transfer, masses and decay widths of the
heavy-quark states cannot be calculated within the framework of
perturbative QCD. As a result, many different approaches have been
developed, for example, based on heavy-quark effective theory
(HQET)~\cite{Mannel199138}, nonrelativistic and relativistic potential
models~\cite{PhysRevLett.48.1653}, or lattice QCD~\cite{Onogi:2009aa}.

In the limit of HQET, heavy-quark mesons, comprised of one heavy and
one light quark, are the closest analogy to the hydrogen atom, which
provided important tests of quantum electrodynamics. Heavy-quark
baryons, comprised of one heavy and two light quarks, extend the
hydrogen atom analogy of HQET by treating the two light quarks as a
diquark system. This leads to degenerate spin-1/2 states resulting
from the combination of a spin-0, or a spin-1, light diquark with the
heavy quark, and thus represents a complementary situation compared to
heavy-quark mesons. Measurements of the mass spectrum and spin
splittings of heavy-quark baryons are important for validating the
theoretical techniques, and build confidence in their predictions for
other heavy flavor studies.

In this paper, we measure the properties of heavy-quark baryons that
contain a $c$ quark, namely the resonances $\Lambda_c(2595)^+$,
$\Lambda_c(2625)^+$, $\Sigma_c(2455)^{++,0}$, and
$\Sigma_c(2520)^{++,0}$. For simplification, we refer to
$\Sigma_c^{++,0}$ as $\Sigma_c$ wherever this information is not
crucial. Throughout the paper, the use of a specific particle state
implies the use of the charge-conjugate state as well. The quark model
predicts the $\Lambda_c(2595)^+$ and $\Lambda_c(2625)^+$, referred to
as $\Lambda_c^{*+}$, to be the lowest orbital excitations of the
$\Lambda_c^+$ groundstate with a spin-0 light diquark. The two
$\Sigma_c$ resonances are expected to have no orbital excitation and a
spin-1 light diquark.

\begin{table}[hb]
  \caption{Theoretical predictions for the masses of the charmed baryons under study. All values are given in \mevcc.}
  \begin{tabular}{cccccc} \hline \hline 
    Hadron & \cite{Mathur:2002ce} & \cite{Roberts:2007ni}  & \cite{Ebert:2007nw,Ebert:2005xj} & \cite{Zhang:2008pm} & \cite{Bernotas:2008bu} \\ \hline
    $\Sigma_c(2455)$ & 2452 & 2455 & 2439 & $2400\pm310$ & 2393 \\
    $\Sigma_c(2520)$ & 2538 & 2519 & 2518 & $2560\pm240$ & 2489 \\
    $\Lambda_c(2595)^+$ & $\cdot\cdot\cdot$ & 2625 & 2598 & $2530\pm220$ & $\cdot\cdot\cdot$ \\
    $\Lambda_c(2625)^+$ & $\cdot\cdot\cdot$ & 2636 & 2628 & $2580\pm240$ & $\cdot\cdot\cdot$ \\
    \hline \hline 
  \end{tabular}
  \label{tab:theoryPredictions}
\end{table}

Some theoretical predictions of the resonance masses are summarized in
Table~\ref{tab:theoryPredictions}, where Ref.~\cite{Mathur:2002ce}
uses lattice QCD,
Refs.~\cite{Roberts:2007ni,Ebert:2007nw,Ebert:2005xj} are based on the
quark model, Ref.~\cite{Zhang:2008pm} employs QCD sum rules and
Ref.~\cite{Bernotas:2008bu} uses a bag model. There are a few
calculations that predict the $\Sigma_c(2455)$ natural width in the
region of
1--3\,\mevcc~\cite{Ivanov:1998qe,Ivanov:1999bk,Tawfiq:1998nk,Huang:1995ke,Pirjol:1997nh,Rosner:1995yu}
and the $\Sigma_c(2520)$ width to be about
18\,\mevcc~\cite{Rosner:1995yu}. No predictions are available for the
$\Lambda_c(2595)^+$ and $\Lambda_c(2625)^+$ widths.

\begin{table}
  \caption{World average values of the mass differences between the charmed baryon resonances and the $\Lambda_c^+$ mass, $\Delta M$, and their natural widths, $\Gamma$~\cite{PDG}.}
  \begin{tabular}{ccc} \hline \hline
    Hadron & $\Delta M$ [\mevcc] & $\Gamma$ [\mevcc] \\ \hline
    $\Sigma_c(2455)^{++}$ & $167.56\pm0.11$ & $2.23\pm0.30$ \\
    $\Sigma_c(2455)^0$ & $167.30\pm0.11$ & $2.2\pm0.4$ \\
    $\Sigma_c(2520)^{++}$ & $231.9\pm0.6$ & $14.9\pm1.9$ \\
    $\Sigma_c(2520)^0$ & $231.6\pm0.5$ & $16.1\pm2.1$ \\
    $\Lambda_c(2595)^+$ & $308.9\pm0.6$ & $3.6^{+2.0}_{-1.3}$ \\
    $\Lambda_c(2625)^+$ & $341.7\pm0.6$ & $<1.9\, \mathrm{at\,\, 90\%\,\, C.L.}$ \\
    \hline \hline
  \end{tabular}
  \label{tab:experimentalInfo}
\end{table}

Experimental observation of all four states studied here and
measurements of some of their properties have been reported
earlier~\cite{Artuso:2001us,Link:2001ee,Brandenburg:1996jc,Athar:2004ni,Albrecht:1997qa,Frabetti:1994,Frabetti:1995sb,Edwards:1994ar}.
We list the world average masses and widths in
Table~\ref{tab:experimentalInfo}, omitting $\Sigma_c^+$ states, which
are difficult to reconstruct with the CDF II detector due to the
inefficiency in $\pi^0$ identification. For $\Sigma_c(2455)$, many
measurements exist with most of the information coming from
CLEO~\cite{Artuso:2001us} and FOCUS~\cite{Link:2001ee}. Experimental
information on the $\Sigma_c(2520)$ states comes exclusively from
CLEO~\cite{Brandenburg:1996jc,Athar:2004ni} and it is worth noting
that the two measurements of the $\Sigma_c(2520)^{++}$ mass are
inconsistent. For $\Lambda_c(2595)^+$ and $\Lambda_c(2625)^+$ three
experiments have contributed, namely ARGUS~\cite{Albrecht:1997qa},
E687 at Fermilab~\cite{Frabetti:1994,Frabetti:1995sb} and
CLEO~\cite{Edwards:1994ar}, all of which suffer from rather small data
samples. In addition, Blechman and co-authors~\cite{Blechman:2003mq}
showed that a more sophisticated treatment of the mass line shape,
which takes into account the proximity of the $\Lambda_c(2595)^+$ mass
to the sum of the masses of its decay products, yields a
$\Lambda_c(2595)^+$ mass which is 2--3\,\mevcc lower than the one
observed. The $\Sigma_c(2455)$ and $\Sigma_c(2520)$ decay directly to
$\Lambda_c^+\pi$, whereas the $\Lambda_c(2595)^+$ and
$\Lambda_c(2625)^+$ end mainly in a $\Lambda_c^+\pi\pi$ final state
with dominating decays through intermediate $\Sigma_c$ resonances.
Therefore, these four resonances contribute to each other's
background, which requires a dedicated cross-feed background modeling
in each case.

In this analysis, we exploit a large sample of $\Lambda_c^+ \to
p\,K^-\,\pi^+$ decays produced in $p\bar{p}$ collisions at $\sqrt{s}$
= 1.96\,TeV and collected by the CDF II detector. Measurements of the
masses and widths of the charmed baryons are performed through fits to
the reconstructed mass distributions calculated from the momenta of
the final state tracks. We take into account all expected cross-feeds
and threshold effects.

The paper is organized as follows. In Sec.~\ref{sec:detector} we
briefly describe the CDF II detector and the trigger components
important for this analysis. Secs.~\ref{sec:reconstruction} and
\ref{sec:selection} describe the candidate reconstruction and
selection, respectively. In Sec.~\ref{sec:fit} we explain the fits
involved in the measurements, followed by a discussion of systematic
uncertainties in Sec.~\ref{sec:systematics}. Finally the results and
conclusions are presented in Sec.~\ref{sec:results}.

\section{CDF II detector and trigger}
\label{sec:detector}

Among the components and capabilities of the CDF II
detector~\cite{Acosta:2004yw}, the tracking system is the one most
relevant to this analysis. It lies within a uniform, axial magnetic
field of $1.4$\,T strength. The inner tracking volume up to a radius
of $28$\,cm is comprised of 6--7 layers of double-sided silicon
microstrip detectors~\cite{Hill:2004qb}. An additional layer of
single-sided silicon is mounted directly on the beam-pipe at a radius
of $1.5$\,cm, allowing excellent resolution on the impact parameter
$d_0$, defined as the distance of closest approach of the track to the
interaction point in the plane transverse to the beam line. The
silicon detector provides a vertex resolution of approximately
$15\,\text{\textmu} \mathrm{m}$ in the transverse and
$70\,\text{\textmu} \mathrm{m}$ in the longitudinal direction. The
remainder of the tracking volume from a radius of $40$ to $137$\,cm is
occupied by an open-cell drift chamber (COT)~\cite{Affolder:2003ep},
providing a transverse momentum resolution of $\sigma(p_T)/p_T^2
\approx 0.1\%/(\mathrm{GeV}/c)$. Hadron identification, which is
crucial for distinguishing slow kaons and protons from pions, is
achieved by a likelihood combination of information from a
time-of-flight system (TOF)~\cite{Acosta:2004kc} and ionization energy
loss in the COT. This offers about $1.5\sigma$ separation between
kaons, or protons, and pions.

A three-level trigger system is used for the online event selection.
The most important device for this analysis at level 1 is the
extremely fast tracker (XFT)~\cite{Thomson:2002xp}. It identifies
charged particles using information from the COT and measures their
transverse momenta and azimuthal angles around the beam direction. The
basic requirement at level 1 is two charged particles with transverse
momentum, $p_T$, greater than 2\,\gevc. At level 2, the silicon vertex
trigger~\cite{Ristori:2010} adds silicon hit information to the XFT
tracks, thus allowing the precise measurement of impact parameters of
tracks. The two level 1 tracks are required to have impact parameters
between $0.1$ and 1\,mm and to be consistent with coming from a common
vertex displaced from the interaction point by at least
100\,\text{\textmu}m in the plane transverse to the beam line. The
level 3 trigger is implemented in software and provides the final
online selection by confirming the first two trigger-level decisions
using a more precise reconstruction similar to the offline software.
This trigger is designed to collect hadronic decays of long-lived
particles such as $b$ and $c$ hadrons. As determined by a study of the
impact parameter distributions, the sample of charmed baryons recorded
by the trigger consists of approximately equal contributions from
$\Lambda_b$ decays and direct $c\overline{c}$ production.

\section{Data set and reconstruction}
\label{sec:reconstruction}

The analysis is performed on a data set collected by the CDF II
detector at the Tevatron $p\bar{p}$ collider between February 2002 and
June 2009 corresponding to an integrated luminosity of
$5.2\,\mathrm{fb^{-1}}$. The data were accumulated using the displaced
two track vertex trigger described in the previous Section.

The offline reconstruction of candidates starts with refitting tracks
using pion, kaon and proton mass hypotheses to properly take into
account differences in the multiple scattering and ionization energy
loss. In the second step, three tracks, one with pion, one with kaon,
and one with proton mass hypotheses, are combined to form a
$\Lambda_c^+$ candidate. The three tracks are subjected to a kinematic
fit that constrains them to originate from a common vertex. We require
that the proton and pion candidates have the same charge and that the
total charge of all three tracks is $\pm1$. To construct
$\Sigma_c(2455)$ and $\Sigma_c(2520)$ candidates we combine each
$\Lambda_c^+$ candidate with one of the remaining tracks in the event
using a pion mass hypothesis. The $\Lambda_c(2595)^+$ and
$\Lambda_c(2625)^+$ candidates are obtained by combining each
$\Lambda_c^+$ candidate with all possible oppositely charged track
pairs taken from the remaining tracks in the event using the pion mass
hypothesis for each of them. The tracks forming each baryon candidate
are subjected to a kinematic fit that constrains them to originate
from a common point. In each step of the reconstruction, standard
quality requirements on tracks and vertices are used to ensure
well-measured masses and decay-positions.

We use simulated events to estimate the detector mass resolutions of
the charmed baryons studied here. The decays are simulated by means of
the {\sc evtgen} package~\cite{Lange:2001}, where the $\Lambda_c^+$ is
forced to decay into $pK^-\pi^+$ with its resonance structure taken
into account. Afterwards, the generated events are passed through the
detector simulation and then reconstructed by the same software used
for data.

\section{Candidate selection}
\label{sec:selection}

\begin{figure}
  \centerline{\includegraphics[width=8cm]{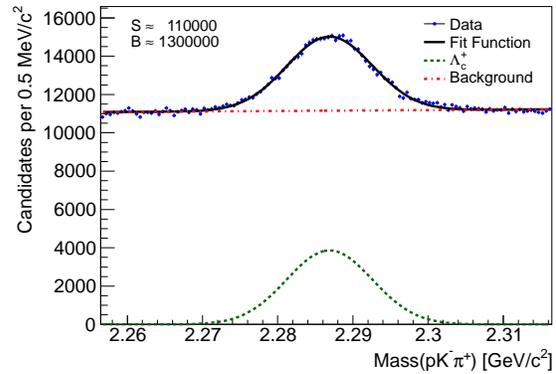}}
  \caption{(color online) The mass distribution of $\Lambda_c^+$
    candidates used to train one of the two neural networks for the
    $\Lambda_c^+$ selection.}
  \label{fig:LcMassprecuts}
\end{figure}

\begin{table}
  \caption{Inputs to the neural network for the $\Lambda_c^+$
    selection sorted by their importance.}
  \begin{tabular}{cccc} \hline \hline
    Index & Variable & Index & Variable  \\ \hline 
    1  & $LL_p(p)$  & 8  &  $p_T(p)$ \\ 
    2  & $\sigma_{L_{xy}}(\Lambda_c^+)$  & 9  & $\cos(\sphericalangle(\Lambda_c^+,K))$  \\ 
    3  & $LL_K(K)$  & 10  & $p_T(\pi)$   \\ 
    4  & $\cos(\sphericalangle(\Lambda_c^+,p))$  & 11  & $d_0/\sigma_{d_0}(K)$  \\ 
    5  & $\chi^2(\Lambda_c^+)$  & 12  & $p_T(K)$  \\ 
    6  & $L_{xy}(\Lambda_c^+)$  & 13  & $d_0/\sigma_{d_0}(p)$  \\ 
    7  & $d_0/\sigma_{d_0}(\pi)$  &   &   \\ \hline \hline
  \end{tabular}
  \label{tab:LcNNInputs}
\end{table}

\begin{figure}
  \centerline{\includegraphics[width=8cm]{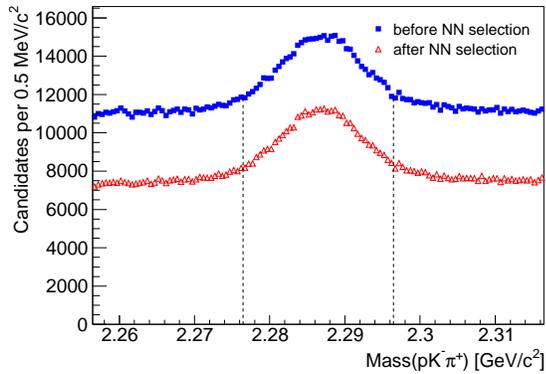}}
  \caption{(color online) The mass distributions of $\Lambda_c^+$
    candidates before (blue full squares) and after (red open
    triangles) requiring their neural network output to correspond to
    an \textit{a posteriori} signal probability greater than $2.5\%$.
    The vertical dashed lines indicate a $\pm 10\,\mevcc$ region
    around the nominal $\Lambda_c^+$ mass~\cite{PDG} used for the
    selection of the $\Sigma_c$ and $\Lambda_c^{*+}$ states.}
  \label{fig:LcMassNN}
\end{figure}

The selection of the candidates is done in two steps. In each one we
first impose some quality requirements to suppress the most obvious
background. For the surviving candidates we use a neural network to
distinguish signal from background. Since all final states feature a
$\Lambda_c^+$ daughter, the first step is the $\Lambda_c^+$ selection.
In the second step, we perform a dedicated selection of the four
states under study. All neural networks are constructed with the
NeuroBayes package~\cite{Feindt:2004aa,Feindt:2006pm} and trained,
only using data, by means of the $_s\mathcal{P}lot$
technique~\cite{Pivk:2004ty,Pivk:2006iy}. This technique assigns a
weight to each candidate proportional to the probability that the
candidate is signal. The candidate weight is based on the
discriminating variables, which are required to be independent of the
ones used in the neural network training. In our case, the
discriminating variable is the invariant mass of the candidate. In the
training, each candidate enters with a weight calculated from the
signal probability that is derived from its mass. Based on these
weights, the neural network can learn the features of signal and
background events. Since we use only data for the neural network
trainings, we randomly split each sample into two parts (even and odd
event numbers) and train two networks. Each of them is then applied to
the complementary subsample in order to maintain a selection which is
trained on a sample independent from the one to which it is applied.
This approach avoids a bias of the selection originating from
statistical fluctuations possibly learnt by the network. Additionally,
using candidates from two different mass regions populated by
background only for the training, we verify that the network selection
does not depend on the mass or create an artificial excess in the
spectrum.

\subsection{$\Lambda_c^+$ selection}

The $\Lambda_c^+ \to p\,K^-\,\pi^+$ candidates are required to decay
to a proton with $p_T > 1.9\,\gevc$ and other particles with $p_T >
400\,\mevc$. The displacement of the associated secondary vertex,
projected onto the $\Lambda_c^+$ transverse momentum direction, to the
beam, $L_{xy}$, is required to be greater than 0.25\,mm. In addition,
we use particle identification information from the TOF and
$\mathrm{d}E/\mathrm{d}x$ from the COT. We combine the two sources of
information for each track $t$ into a single variable
\begin{equation}
  LL_i(t)=\frac{P_{\mathrm{d}E/\mathrm{d}x}^i(t)P_{TOF}^i(t)}
   {\sum_{j=\pi,K,p} f_jP_{\mathrm{d}E/\mathrm{d}x}^j(t)P_{TOF}^j(t) },
\end{equation}
where the index $i$ denotes the hypothesis of the particle type. The
$P_{TOF}^i(t)$ is the probability to observe the measured
time-of-flight given a particle of type $i$, and
$P_{\mathrm{d}E/\mathrm{d}x}^i(t)$ is the probability to observe the
measured $\mathrm{d}E/\mathrm{d}x$. The fractions $f_j$ are
$f_\pi=0.7$, $f_K=0.2$, and $f_p=0.1$, as estimated from TOF
information of a generic background sample. We apply the requirement
$LL_p>0.6$ on the proton track and $LL_K>0.2$ on the kaon track. In
case TOF or $\mathrm{d}E/\mathrm{d}x$ information is not available for
a given track, we do not impose the corresponding requirement. The
mass distribution of the candidates with even event numbers is shown
in Fig.~\ref{fig:LcMassprecuts}. A fit with a Gaussian signal and a
linear background function defines the probability density functions
(PDFs) used to calculate the $_s\mathcal{P}lot$ weights for the
$\Lambda_c^+$ network training. The corresponding distribution of
odd-numbered events is similar.

The full list of input quantities of the neural network, sorted by
their importance, can be found in Table~\ref{tab:LcNNInputs}. In the
table, $d_0$ denotes the impact parameter with respect to the primary
vertex of the $p\bar{p}$ interaction for a track in the plane
transverse to the beam direction, $\sigma_{d_0}$ its uncertainty,
$\chi^2(\Lambda_c^+)$ the quality of the kinematic fit of the
$\Lambda_c^+$ candidate, and $\cos(\sphericalangle(\Lambda_c^+,t))$
the cosine of the angle between the momentum of the $\Lambda_c^+$
candidate in the lab frame and the momentum of the proton or kaon
track in the $\Lambda_c^+$ rest frame. These angles carry information
about the resonant substructure of the decay $\Lambda_c^+ \to
p\,K^-\,\pi^+$.

To demonstrate the ability of the neural network to classify signal
and background, the mass distributions of $\Lambda_c^+$ candidates
with even event numbers before and after requiring their neural
network output to correspond to an \textit{a posteriori} signal
probability greater than $2.5\%$ is shown in Fig.~\ref{fig:LcMassNN}.
This requirement leads to a background reduction of $32\%$ while
keeping $97\%$ of the signal. We use the output of the $\Lambda_c^+$
neural network as input to the neural networks for selecting the
$\Sigma_c$ and $\Lambda_c^{*+}$ resonances.

\subsection{$\Sigma_c(2455)$ and $\Sigma_c(2520)$ selection}

\begin{figure*}
  \centerline{\includegraphics[width=8cm]{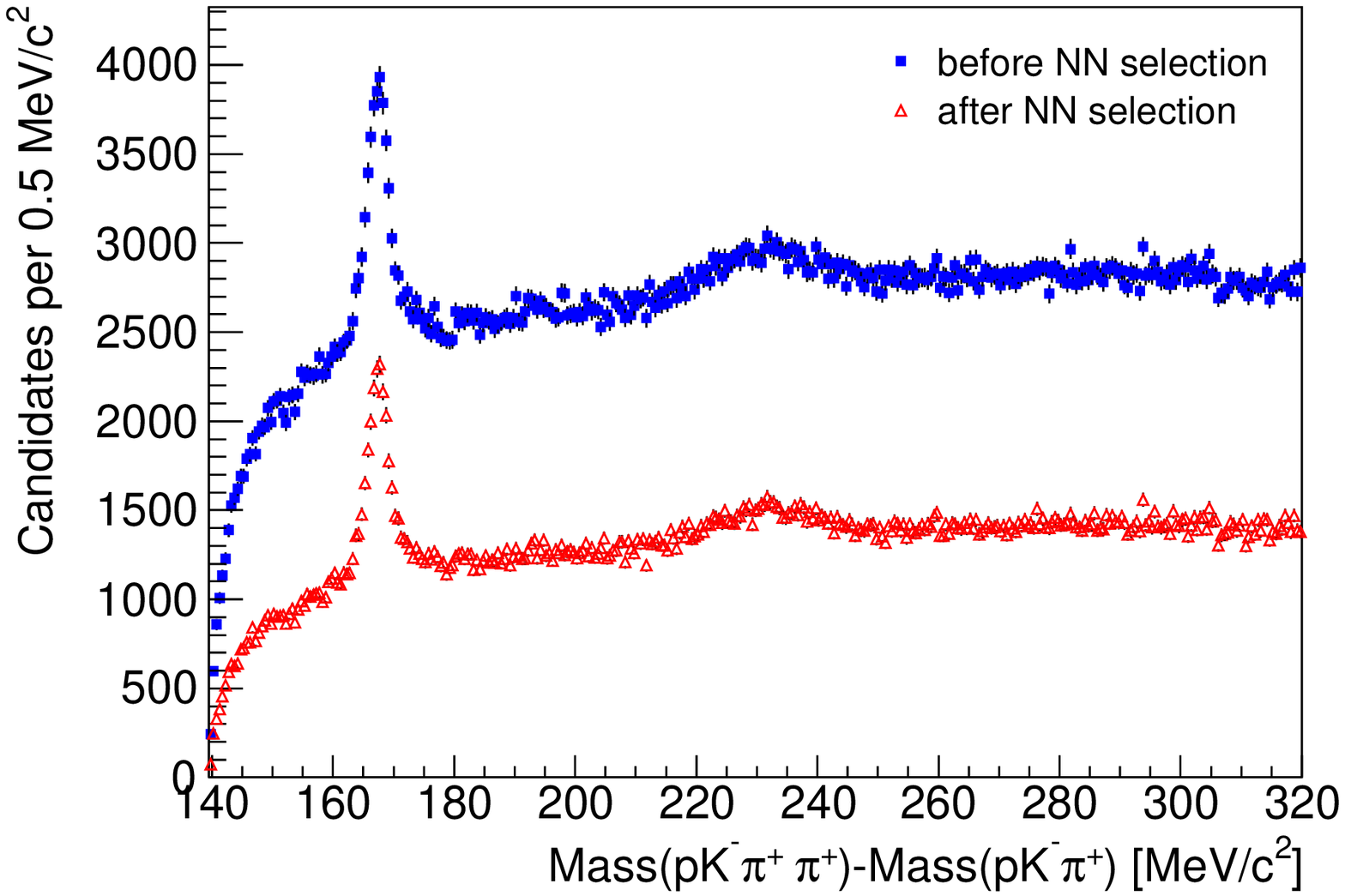}
    \includegraphics[width=8cm]{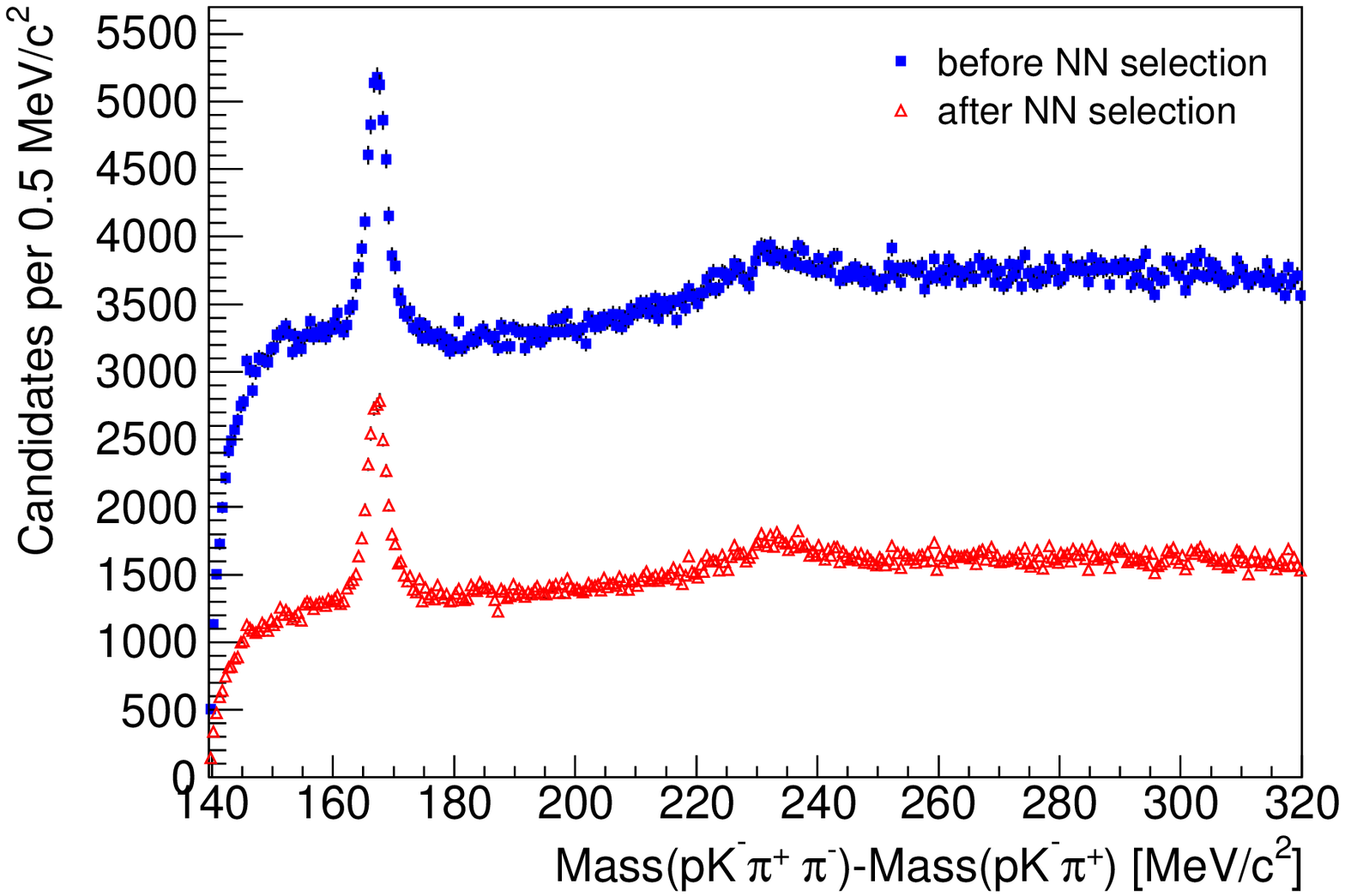}}
  \caption{(color online) The mass difference distributions of the
    $\Lambda_c^+\pi^+$ (left) and $\Lambda_c^+\pi^-$ (right)
    candidates before (blue full squares) and after (red open
    triangles) applying the neural network selection.}
  \label{fig:ScMass}
\end{figure*}

\begin{figure}
  \centerline{\includegraphics[width=8cm]{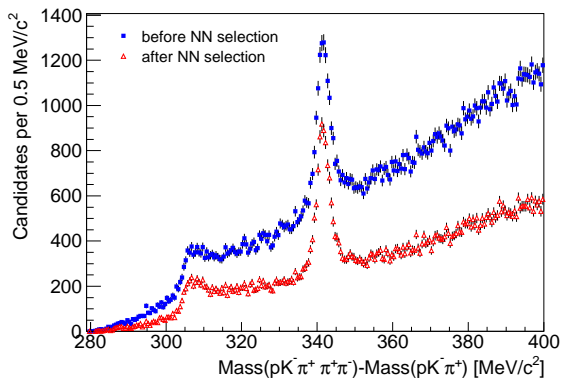}}
  \caption{(color online) The mass difference distribution of the
    $\Lambda_c^+\pi^+\pi^-$ candidates before (blue full squares) and
    after (red open triangles) applying the neural network selection.}
  \label{fig:LcSMass}
\end{figure}

The $\Sigma_c(2455)^{++,0}$ and $\Sigma_c(2520)^{++,0} \to
\Lambda_c^+\,\pi^{+,-}$ selection starts with the application of a few
loose requirements to remove the most obvious background, followed by
the use of a neural network. We require the \textit{a posteriori}
signal probability of the $\Lambda_c^+$ neural network to be greater
than $2.5\%$ (see Fig.~\ref{fig:LcMassNN}), the $p_T(\pi)$ of the
added pion to be greater than 400\,\mevc, $d_0(\pi)<1.5$\,mm, and the
mass of the $\Lambda_c^+$ candidate to be within $\pm 10\,\mevcc$ of
the nominal $\Lambda_c^+$ mass~\cite{PDG}, $2276.46 < M(pK^-\pi^+) <
2296.46\,\mevcc$ (see Fig.~\ref{fig:LcMassNN}). These requirements are
common for both neutral and doubly-charged states. The mass difference
$\Delta M=M(\Sigma_c)-M(\Lambda_c^+)$ distributions of all the
$\Lambda_c^+\pi^+$ and $\Lambda_c^+\pi^-$ candidates are shown in
Fig.~\ref{fig:ScMass}. In the $\Delta M$ definition, $M(\Sigma_c)$ and
$M(\Lambda_c^+)$ correspond to the reconstructed masses of the
$\Sigma_c$ and $\Lambda_c^+$ candidates.

The neural network for the final selection of the $\Sigma_c(2455)$ and
$\Sigma_c(2520)$ candidates uses five input quantities. Ordered by
their importance, these are the output of the $\Lambda_c^+$ neural
network $NN(\Lambda_c^+)$, the proper decay time of the $\Sigma_c$
candidate $t(\Sigma_c) = (L_{xy}(\Sigma_c) \cdot M(\Sigma_c))/(c \cdot
p_T(\Sigma_c))$, the quality of the kinematic fit of the $\Sigma_c$
candidate $\chi^2(\Sigma_c)$, the uncertainty of the $\Sigma_c$ impact
parameter in the transverse plane $\sigma_{d_0}(\Sigma_c)$, and the
impact parameter in the transverse plane of the pion from the
$\Sigma_c$ decay $d_0(\pi)$. Independent neural networks are employed
for $\Sigma_c^{++}$ and $\Sigma_c^0$. The training itself is performed
using candidates in the mass difference region from 155 to
180\,\mevcc. Although this includes only $\Sigma_c(2455)$ candidates,
it is applied to select $\Sigma_c(2520)$ candidates as well. The
$_s\mathcal{P}lot$ weights are determined by a fit to the $\Delta M$
distribution with a Gaussian function for the signal and a linear
function for the background PDF. We choose the threshold on the output
of the $\Sigma_c$ neural network to maximize $S/\sqrt{S+B}$, where $S$
is the number of signal $\Sigma_c$ events and $B$ is the number of
background events in $\Delta M$ between 162.3 and 172.3\,\mevcc. The
$S$ and $B$ yields are derived from a fit to the $\Delta M$
distribution which uses a Gaussian function for the signal and a
linear function for the background and covers the $\Delta M$ range
used for the neural network training. The resulting neural network
output requirement is the same for both charge combinations and
corresponds to an \textit{a posteriori} signal probability of the
neural networks greater than $10\%$. The $\Delta M$ distributions of
the selected candidates are shown in Fig.~\ref{fig:ScMass}.

\subsection{$\Lambda_c(2595)^+$ and $\Lambda_c(2625)^+$ selection}

The initial step of the $\Lambda_c(2595)^+$ and $\Lambda_c(2625)^+ \to
\Lambda_c^+\,\pi^+\,\pi^-$ selection requires the \textit{a
  posteriori} signal probability of the $\Lambda_c^+$ neural network
to be greater than $2.5\%$, $2276.46 < M(pK^-\pi^+) < 2296.46\,\mevcc$
(see Fig.~\ref{fig:LcMassNN}), $p_T(\pi)$ of both added pions to be
greater than 400\,\mevc, and the impact parameter of the object
constructed from the two additional pions to be
$d_0(\pi^+\pi^-)<1.0$\,mm. The mass difference $\Delta
M=M(\Lambda_c^{*+})-M(\Lambda_c^+)$ distribution is shown in
Fig.~\ref{fig:LcSMass}.

We use the $\Delta M$ region between 327 and 357\,\mevcc for the
neural network training. Although this includes only
$\Lambda_c(2625)^+$ candidates, it is applied to select
$\Lambda_c(2595)^+$ candidates as well. The $_s\mathcal{P}lot$ weights
are based on a fit to the $\Delta M$ distribution with a Gaussian
function for the signal and a linear function for the background PDF.
The neural network uses four inputs. Ordered by their importance,
these are the quality of the $\Lambda_c^{*+}$ kinematic fit
$\chi^2(\Lambda_c^{*+})$, the uncertainty of the impact parameter of
the combined two-pion object $\sigma_{d_0}(\pi^+\pi^-)$, the output of
the $\Lambda_c^+$ neural network $NN(\Lambda_c^+)$, and the proper
decay time of the $\Lambda_c^{*+}$ candidate $t(\Lambda_c^{*+})$. We
choose the requirement that maximizes $S/\sqrt{S+B}$, corresponding to
an \textit{a posteriori} signal probability of the neural network
greater than $12.5\%$. The $S$ and $B$ yields are derived from a fit
to the $\Delta M$ distribution using a Gaussian function for the
signal and a linear function for the background, where we consider
events in the region $336.7 < \Delta M < 346.7\,\mevcc$. The resulting
mass difference distribution after the final requirements is shown in
Fig.~\ref{fig:LcSMass}.

\section{Fit description}
\label{sec:fit}

To determine the mass differences relative to the $\Lambda_c^+$ and
the widths of the six studied states, we perform binned maximum
likelihood fits to three separate mass difference distributions. The
first two are $\Lambda_c^+\pi^+$ and $\Lambda_c^+\pi^-$, where the
states $\Sigma_c(2455)^{++,0}$ and $\Sigma_c(2520)^{++,0}$ are
studied. The last one is $\Lambda_c^+\pi^+\pi^-$ for
$\Lambda_c(2595)^+$ and $\Lambda_c(2625)^+$. In the case of the
$\Sigma_c$ states, part of the background comes from $\Lambda_c^{*+}$
decays and thus has different properties compared to the combinatorial
background. On the other hand, when fitting $\Lambda_c^{*+}$ states,
there is a background contribution from random
$\Sigma_c^{++,0}\pi^{-,+}$ combinations which have a threshold close
to the $\Lambda_c(2595)^+$ state.

The negative logarithm of the likelihood function has a general form
of
\begin{equation}
 \begin{split}
   - \ln \mathcal{L}(\vec{a}) & = - \sum_{j=1}^{J} \ln
   \left(\frac{\mu_{j}^{n_j} e^{- \mu_j}}{n_j!}\right)\\
   & = - \sum_{j=1}^{J} n_j \ln \mu_j + \sum_{j=1}^{J} \mu_j +
   \sum_{j=1}^{J} \ln (n_j!),
 \end{split}
  \label{eq:Likelihood}
\end{equation}
where $\vec{a}$ are the free parameters, $J$ is the number of bins in
the histogram of the corresponding mass difference distribution, $n_j$
is the number of entries in bin $j$, and $\mu_j$ is the expected
number of entries in bin $j$. The values $\mu_j$ are obtained using
the function
\begin{equation}
  \mu(\Delta M) = N_1 \cdot s_1(\Delta M) + N_2 \cdot
  s_2(\Delta M) + b(\Delta M),
  \label{eq:Sc_Fct}
\end{equation}
where $s_1(\Delta M)$ and $s_2(\Delta M)$ are the PDFs for the two
signals, $b(\Delta M)$ is the background function and $N_i$ are the
corresponding numbers of events. All three PDFs depend on a subset of
the free parameters $\vec{a}$. The function is evaluated at the bin
center to calculate the expectation for $\mu_j$. While the general
structure is the same in all three fits, the PDFs are specific to
$\Sigma_c$ and $\Lambda_c^{*+}$ states.

\subsection{$\Sigma_c(2455)$ and $\Sigma_c(2520)$ fit}

\begin{figure}
  \centering
  \includegraphics[width=8cm]{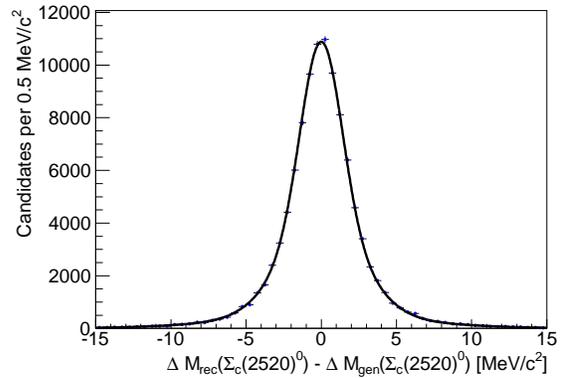}
  \caption {$\Sigma_c(2520)^0$ mass resolution obtained from simulated
    events by subtracting the generated mass difference $\Delta
    M_{\mathrm{gen}}(\Sigma_c(2520)^0)$ from the reconstructed one
    $\Delta M_{\mathrm{rec}}(\Sigma_c(2520)^0)$. The fitted function
    is a combination of three Gaussians with mean zero.}
  \label{fig:Sc02_ResRew.eps}
\end{figure}

In each of the two distributions we need to parametrize two signals
and several background components. We use a 150--320\,\mevcc range to
avoid complications arising from the description of the steep rise of
the background at threshold. Both $\Sigma_c(2455)$ and
$\Sigma_c(2520)$ are described by a nonrelativistic Breit-Wigner
function,
\begin{equation}
  \frac{\mathrm{d}N}{\mathrm{d}\Delta M} \propto \frac{\Gamma}{(\Delta M - \Delta M_0)^2 + \Gamma^2/4},
  \label{eq:BW}
\end{equation}
convolved with a resolution function. The resolution function itself
is parametrized by three Gaussians with mean zero and the other
parameters derived from simulated events. The average width of the
resolution function is about 1.6\,\mevcc for $\Sigma_c(2455)^{++,0}$
and about 2.6\,\mevcc for $\Sigma_c(2520)^{++,0}$. For illustration,
the simulated $\Sigma_c(2520)^0$ mass resolution is shown in
Fig.~\ref{fig:Sc02_ResRew.eps}.

We introduce a single common scaling factor $s$ for the widths of all
three Gaussians to correct for a possible mismatch in our mass
resolution estimate. This scaling factor is allowed to float within a
Gaussian constraint in the fit, what corresponds to adding
\begin{equation}
  0.5 \cdot \left(\frac{s - \mu}{\sigma}\right)^2
  \label{eq:GC}
\end{equation}
with $\mu=1$ and $\sigma=0.2$, reflecting a 20\% uncertainty on the
mass resolution (see Sec.~\ref{sec:systematics}), to the negative
logarithm of the likelihood.

\begin{figure*}
  \centerline{\includegraphics[width=8cm]{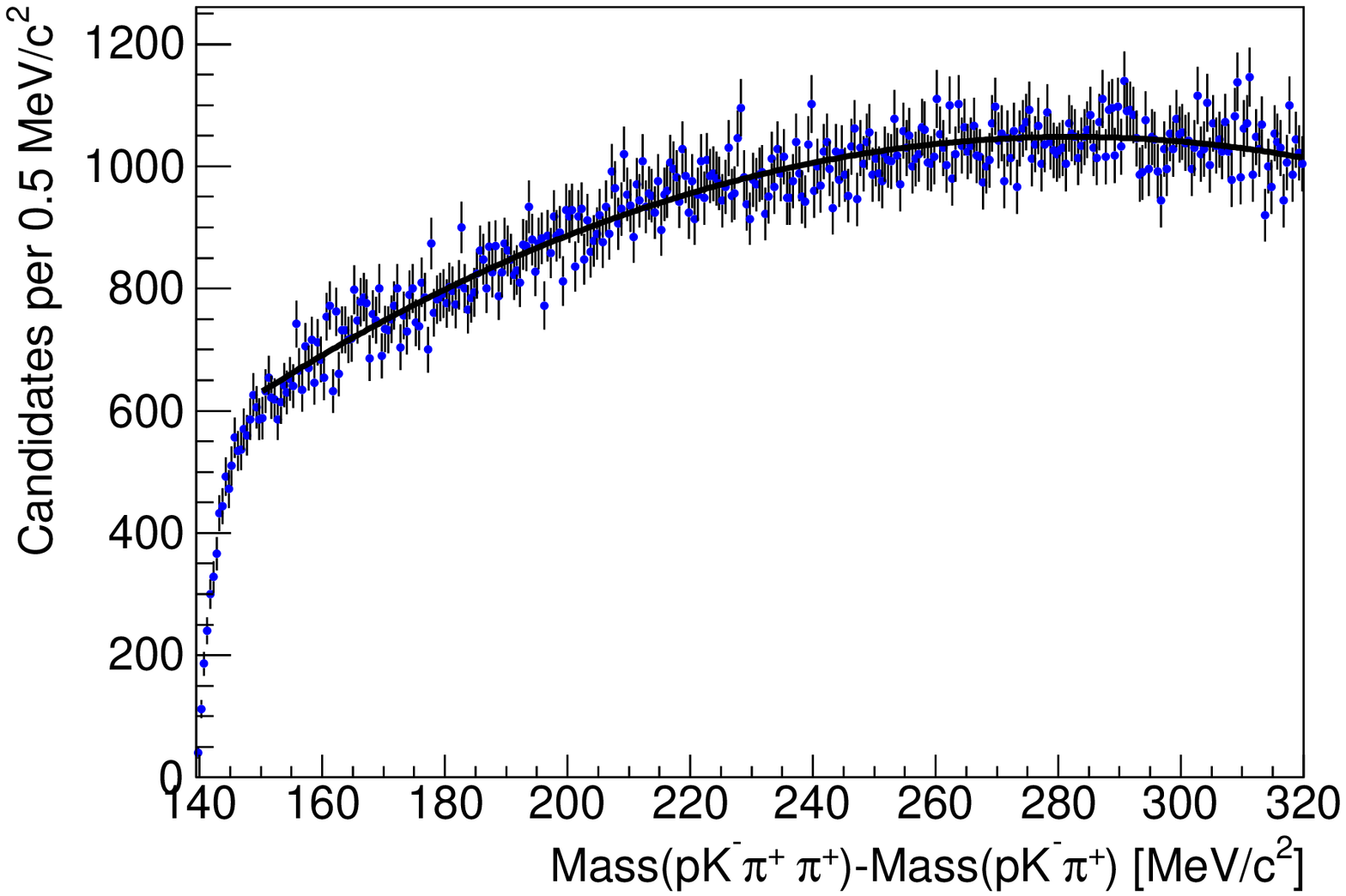}
    \includegraphics[width=8cm]{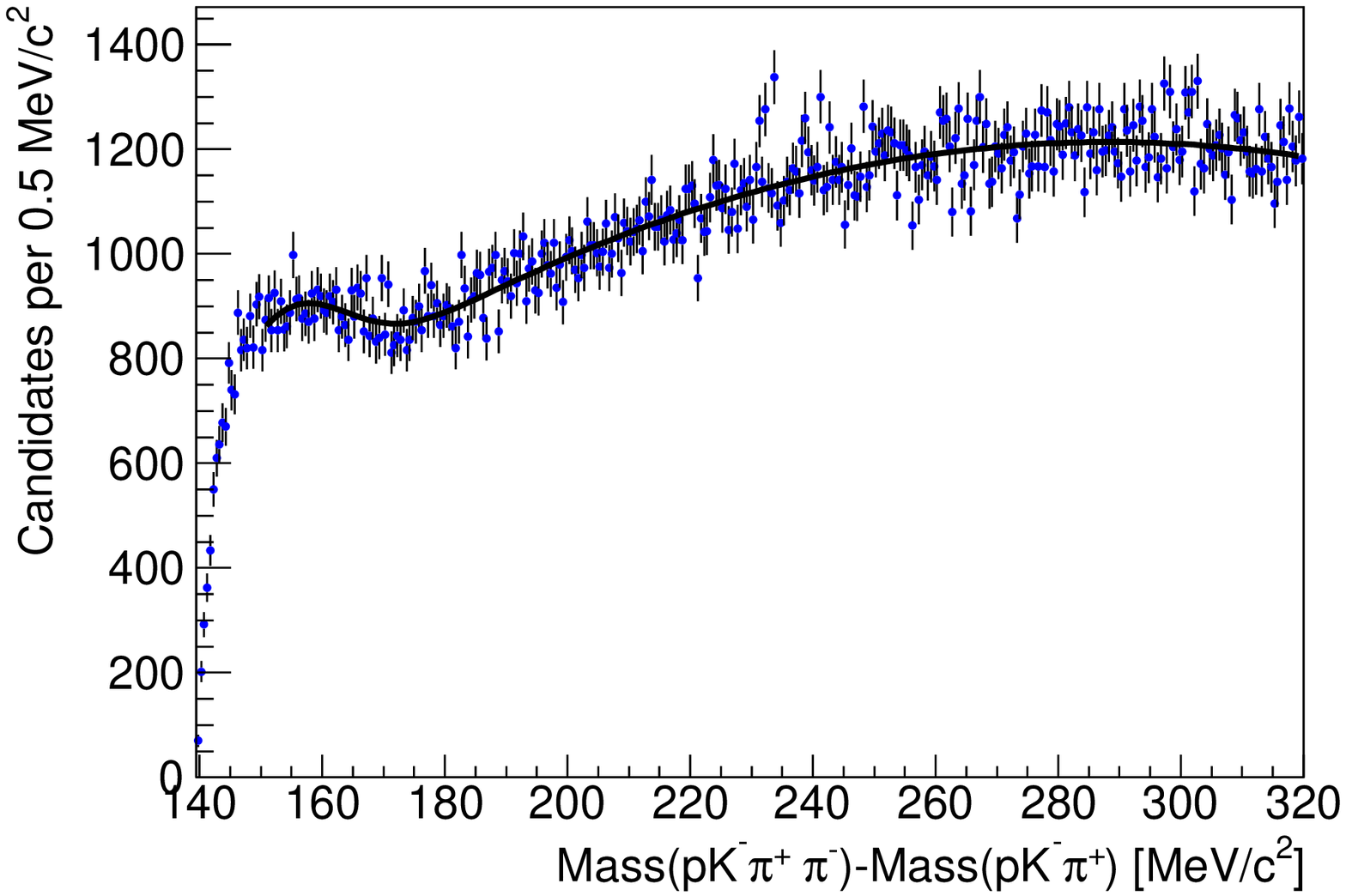}}
  \caption{Fit to the $M(pK^-\pi^+\,\pi^+)-M(pK^-\pi^+)$ (left) and
    $M(pK^-\pi^+\,\pi^-)-M(pK^-\pi^+)$ (right) distributions of the
    candidates from $\Lambda_c^+$ mass sidebands.}
  \label{fig:ScCombinatorial}
\end{figure*}

Three different types of background are considered, namely, random
combinations without real $\Lambda_c^+$, combinations of real
$\Lambda_c^+$ with a random pion, and events due to the decay of
$\Lambda_c^{*+}$ to $\Lambda_c^+\pi^+\pi^-$. The random combinations
without a real $\Lambda_c^+$ dominate and are described by a
second-order polynomial with shape and normalization derived in a fit
to the $\Delta M$ distribution from the $\Lambda_c^+$ mass sidebands
$2261.46 < M(pK^-\pi^+) < 2266.46\,\mevcc$ and $2306.46 < M(pK^-\pi^+)
< 2311.46\,\mevcc$. In the $\Sigma_c$ fit, this contribution is
allowed to float within a Gaussian constraint implemented by the
addition of
\begin{equation}
  0.5 \cdot \vec{\Delta}^T \cdot \mathrm{\bf{V}}^{-1} \cdot \vec{\Delta}
  \label{eq:GCMatrix}
\end{equation}
to the negative logarithm of the likelihood, where \begin{bf}V\end{bf}
is the covariance matrix of the fit to the $\Delta M$ distribution
from the $\Lambda_c^+$ mass sidebands and $\vec{\Delta}$ is the vector
of parameters of the second-order polynomial. The fits to the
distributions from the $\Lambda_c^+$ mass sidebands can be found in
Fig.~\ref{fig:ScCombinatorial}. The difference between doubly-charged
and neutral spectra is due to $D^*(2010)^+ \to D^0\,\pi^+$ mesons with
multibody $D^0$ decays, where not all $D^0$ decay products are
reconstructed. In order to describe this reflection, an additional
Gaussian function is used. The second background source consisting of
real $\Lambda_c^+$ combined with a random pion is modeled by a
third-order polynomial, where all parameters are left free in the fit.
The background originating from $\Lambda_c^{*+}$ decays is described
using theoretical considerations. With good approximation, there are
two states that contribute, namely $\Lambda_c(2595)^+$ and
$\Lambda_c(2625)^+$, decaying into a $\Lambda_c^+\pi^+\pi^-$ final
state. The $\Lambda_c(2595)^+$ decays dominantly to a $\Sigma_c\pi$
final state~\cite{PDG} and thus contributes mainly to the signal. We
therefore neglect its contributions to the backgrounds in the
$\Lambda_c\pi$ distributions. On the other hand, the
$\Lambda_c(2625)^+$ decay is dominantly nonresonant~\cite{PDG}. To
model it, we start from a flat $\Lambda_c^+\pi^+\pi^-$ Dalitz plot and
project it on the appropriate axis. Since the shape of the projection
depends on the reconstructed $\Lambda_c(2625)^+ \to
\Lambda_c^+\,\pi^+\,\pi^-$ mass, we use ten different values of this
mass and weight their contribution according to the
$\Lambda_c(2625)^+$ shape we obtain from our fit to the
$\Lambda_c^+\pi^+\pi^-$ data. This contribution amounts to about $2\%$
of the total background.

\begin{figure*}
  \centerline{\includegraphics[width=8cm]{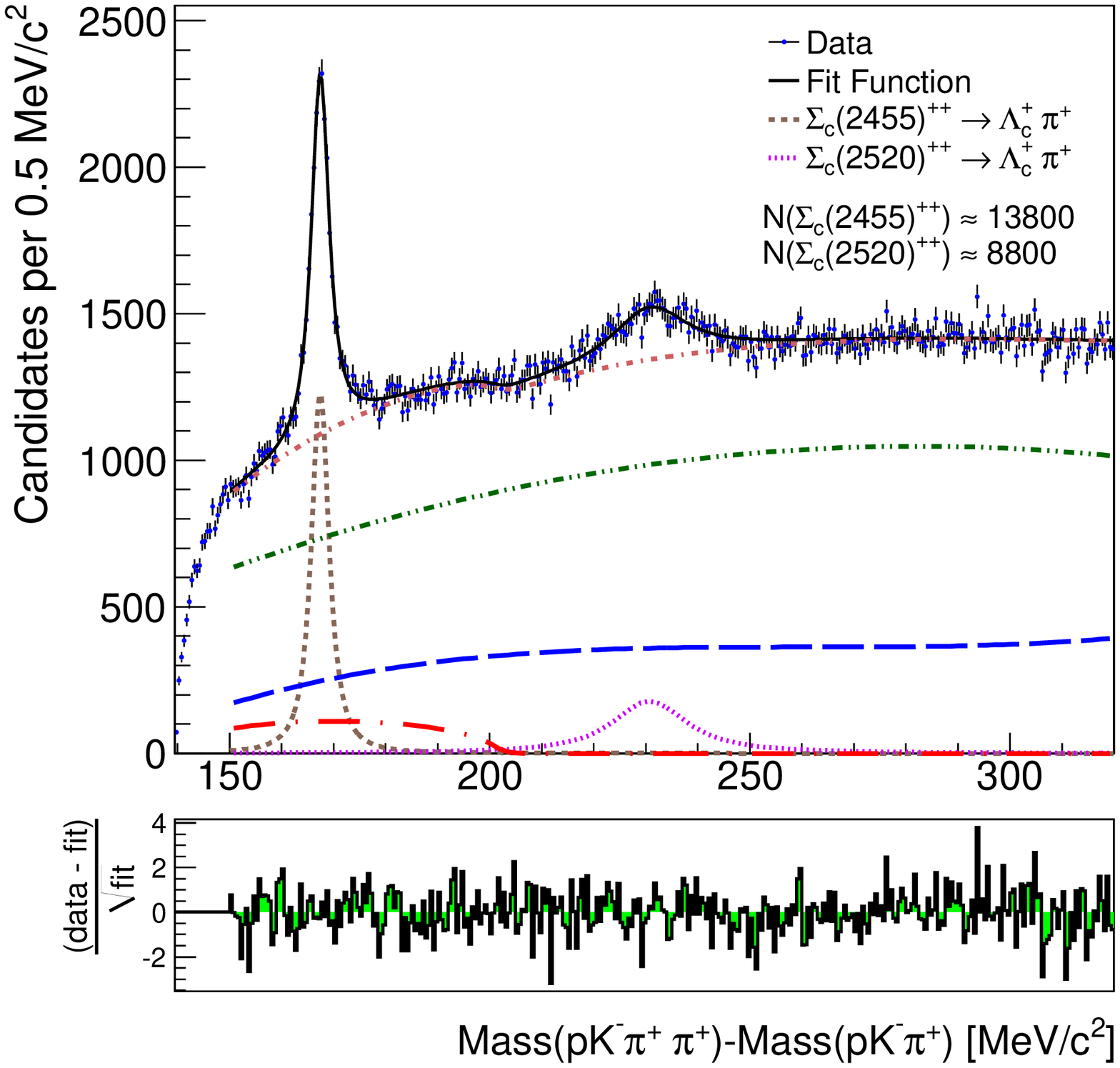}
    \includegraphics[width=8cm]{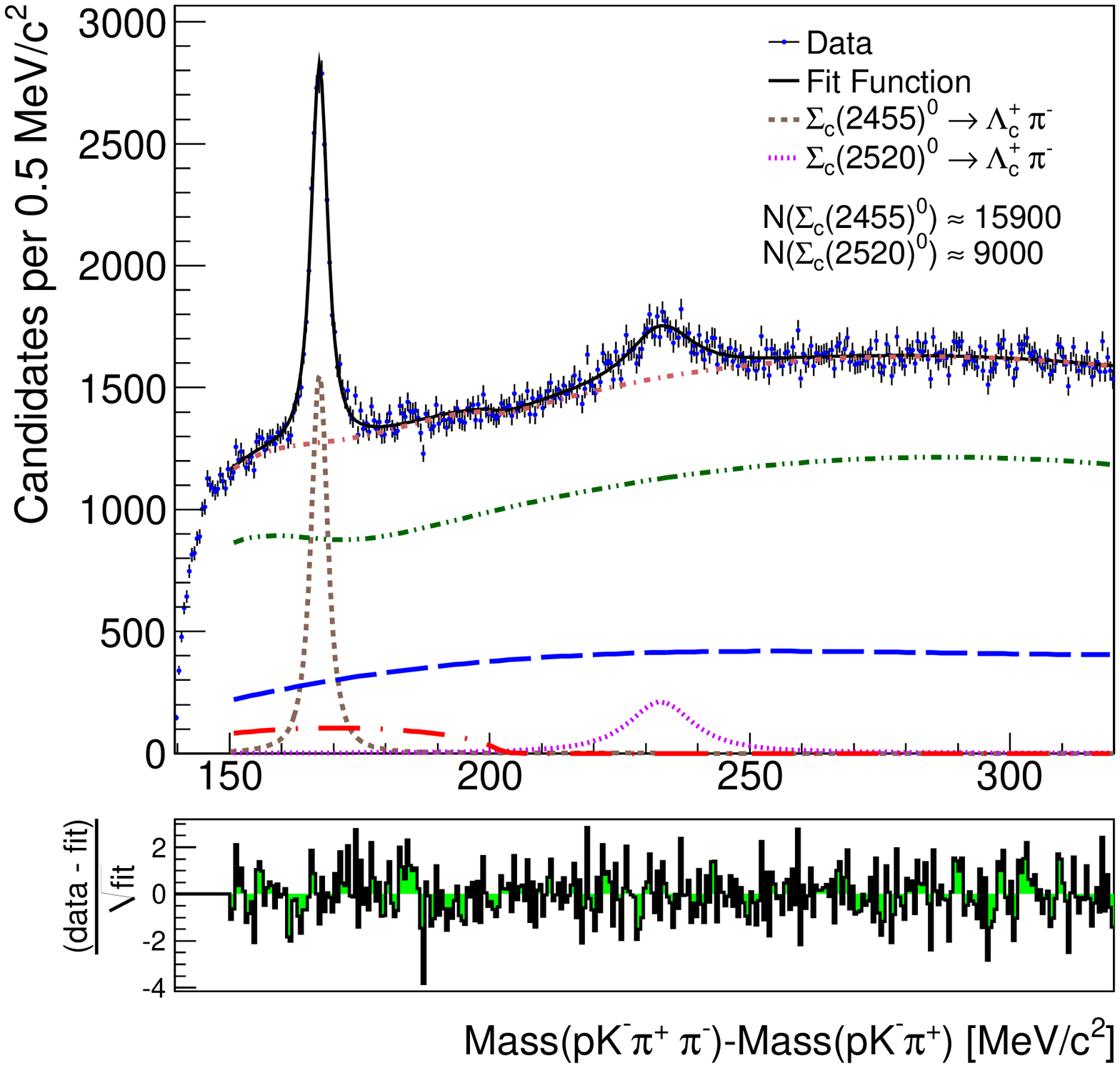}}
  \caption{(color online) The $M(pK^-\pi^+\,\pi^+)-M(pK^-\pi^+)$
    (left) and $M(pK^-\pi^+\,\pi^-)-M(pK^-\pi^+)$ (right)
    distributions obtained from data (points with error bars) together
    with the fits (black solid line). The brown dashed and purple
    dotted lines correspond to the two signal contributions, the green
    dash-double-dotted line represents the combinatorial background
    without real $\Lambda_c^+$, the blue long-dashed line shows real
    $\Lambda_c^+$ combined with a random pion and the red
    long-dash-dotted line represents a reflection from
    $\Lambda_c^{*+}$ decays. The red dash-dotted line corresponds to
    the sum of all three background contributions.}
  \label{fig:Sc_Fit}
\end{figure*}

The full fit to the $\Delta M$ distribution, containing all signal and
background components, can be found in Fig.~\ref{fig:Sc_Fit}. The
$\chi^2$ value of the $\Sigma_c^{++}$ fit is 340 (324 degrees of
freedom) and that of the $\Sigma_c^0$ fit is 384 (321 degrees of
freedom).

\subsection{$\Lambda_c(2595)^+$ and $\Lambda_c(2625)^+$ fit}

The fit for $\Lambda_c(2595)^+$ and $\Lambda_c(2625)^+$ includes two
signals and several background components and is performed in a
$\Delta M$ region from 290 to 400\,\mevcc. Previous measurements of
the $\Lambda_c(2595)^+$ properties indicate that it decays dominantly
to the final state $\Sigma_c\pi$, with the threshold very close to the
$\Lambda_c(2595)^+$ mass~\cite{PDG}. This introduces an additional
complication to the fit compared to the $\Sigma_c$ case. Blechman
\textit{et al.}~\cite{Blechman:2003mq} showed that taking into account
the mass dependence of the natural width yields a lower
$\Lambda_c(2595)^+$ mass measurement than observed by previous
experiments. With the present event sample we are more sensitive to
the details of the $\Lambda_c(2595)^+$ line shape than previous
analyses and include this dependence in the model.

\begin{figure}
  \centering
  \includegraphics[width=8cm]{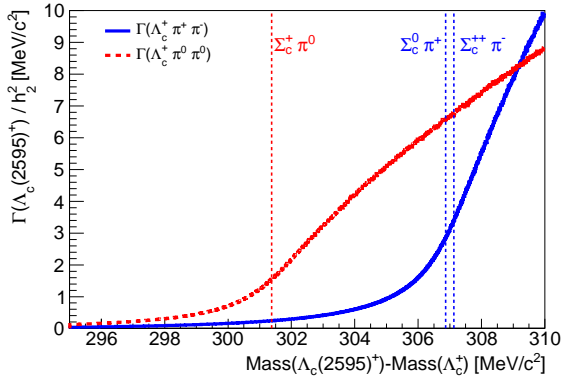}
  \caption{(color online) Calculated dependence of
    $\Gamma(\Lambda_c^+\pi^+\pi^-)$ and
    $\Gamma(\Lambda_c^+\pi^0\pi^0)$ on
    $M(\Lambda_c(2595)^+)-M(\Lambda_c^+)$. The constant factor $h_2^2$
    is determined by a fit to the experimental data.}
\label{fig:Lc(2595)_W}
\end{figure}

The $\Lambda_c(2595)^+$ parametrization follows
Ref.~\cite{Blechman:2003mq}. The state is described by a
nonrelativistic Breit-Wigner function of the form
\begin{widetext}
\begin{equation}
  \frac{\mathrm{d}N}{\mathrm{d}\Delta M} \propto \frac{\Gamma(\Lambda_c^+\pi^+\pi^-)}{(\Delta M - \Delta M_{\Lambda_c(2595)^+})^2 + (\Gamma(\Lambda_c^+\pi^+\pi^-) + \Gamma(\Lambda_c^+\pi^0\pi^0))^2/4}\,,
  \label{eq:Lc(2595)_BW}
\end{equation}
\end{widetext}
where $\Gamma(\Lambda_c^+\pi^+\pi^-)$ and
$\Gamma(\Lambda_c^+\pi^0\pi^0)$ are the mass-dependent partial widths
to the $\Lambda_c^+\pi^+\pi^-$ and $\Lambda_c^+\pi^0\pi^0$ final
states. Assuming that those two final states saturate nearly 100\% of
the $\Lambda_c(2595)^+$ decay width, the sum in the denominator
corresponds to the total width. The two partial widths are derived in
Ref.~\cite{Cho:1994vg} as
\begin{equation}
\begin{split}
  \Gamma&(\Lambda_c^+\pi^+\pi^-) = \frac{g_2^2}{16 \pi^3 f_{\pi}^4} m_{\Lambda_c^+} \int \mathrm{d}E_1 \mathrm{d}E_2 (|\vec{p}_2|^2 |A(E_1)|^2\\
  & + |\vec{p}_1|^2 |B(E_2)|^2 + 2 \vec{p}_1 \cdot \vec{p}_2 \mathrm{Re} [A(E_1) B^*(E_2)]),
\end{split}
  \label{eq:Lc(2595)_Gamma1}
\end{equation}
\begin{equation}
\begin{split}
  \Gamma&(\Lambda_c^+\pi^0\pi^0) = \frac{g_2^2}{16 \pi^3 f_{\pi}^4} m_{\Lambda_c^+} \int \mathrm{d}E_1 \mathrm{d}E_2 (|\vec{p}_2|^2 |C(E_1)|^2\\
  & + |\vec{p}_1|^2 |C(E_2)|^2 + 2 \vec{p}_1 \cdot \vec{p}_2 \mathrm{Re} [C(E_1) C^*(E_2)]).
\end{split}
  \label{eq:Lc(2595)_Gamma2}
\end{equation}
Here, $f_{\pi} = 132\,\mevcc$ is the pion decay
constant~\cite{Follana:2007uv}, $m_{\Lambda_c^+}$ is the world average
$\Lambda_c^+$ mass, $E_1$,$E_2$ are the energies of the two pions in
the rest frame of the $\Lambda_c(2595)^+$, and $\vec{p}_1$,$\vec{p}_2$
are the corresponding momenta. Following Ref.~\cite{Blechman:2003mq},
the coupling constant $g_2$ is determined by the $\Sigma_c$ decay
width using the relation
\begin{equation}
  \Gamma_{\Sigma_c} = \frac{g_2^2}{2 \pi f_{\pi}^2}
  \frac{m_{\Lambda_c^+}}{m_{\Sigma_c}} |\vec{p}_{\pi}|^3,
  \label{eq:Lc(2595)_SigmaGamma}
\end{equation}
with $m_{\Sigma_c}$ being the world average mass of the
$\Sigma_c(2455)$ and $\vec{p}_{\pi}$ the momentum of the pion from the
$\Sigma_c(2455)$ decay to $\Lambda_c\pi$ in the $\Sigma_c(2455)$ rest
frame. From the world average
$\Gamma_{\Sigma_c}=2.2$\,\mevcc~\cite{PDG} we obtain the value $g_2^2
= 0.365$ which is fixed in the fit. The amplitudes $A$, $B$, and $C$
for the decays $\Lambda_c(2595)^+ \to \Sigma_c(2455)^0\,\pi^+$,
$\Lambda_c(2595)^+ \to \Sigma_c(2455)^{++}\,\pi^-$, and
$\Lambda_c(2595)^+ \to \Sigma_c(2455)^+\,\pi^0$ are parametrized as
\begin{equation}
\begin{split}
  A(E) & = \frac{h_2 E}{\Delta m - \Delta m_{\Sigma_c^0} - E + i
    \Gamma_{\Sigma_c^0}/2},
\end{split}
  \label{eq:Lc(2595)_A}
\end{equation}
\begin{equation}
\begin{split}
  B(E) & = \frac{h_2 E}{\Delta m - \Delta m_{\Sigma_c^{++}} - E + i
    \Gamma_{\Sigma_c^{++}}/2},
\end{split}
  \label{eq:Lc(2595)_B}
\end{equation}
\begin{equation}
\begin{split}
  C(E) & = \frac{1}{2} \cdot \frac{h_2 E}{\Delta m - \Delta
    m_{\Sigma_c^+} - E + i \Gamma_{\Sigma_c^+}/2}.
\end{split}
  \label{eq:Lc(2595)_C}
\end{equation}
In these definitions, $m_{\Sigma_c^{++,+,0}}$ and
$\Gamma_{\Sigma_c^{++,+,0}}$ are the mass and the width of the
$\Sigma_c(2455)^{++,+,0}$ taken from Ref.~\cite{PDG}. The coupling
constant $h_2$, defined in Ref.~\cite{Pirjol:1997nh}, is related to
the decay width of the $\Lambda_c(2595)^+$ and represents the actual
quantity we measure instead of the natural width. This approach
describes a purely $S$-wave decay, a possible $D$-wave contribution is
assumed to be negligible and ignored. For illustration, we show the
dependence of the two partial widths on $M(\Lambda_c(2595)^+) -
M(\Lambda_c^+)$ in Fig.~\ref{fig:Lc(2595)_W}. The shape defined by
Eq.~\ref{eq:Lc(2595)_BW} is then numerically convolved with a
resolution function determined from simulation and consisting of three
Gaussians with mean zero. The average width of the resolution function
is about 1.8\,\mevcc. As for the $\Sigma_c$ case, we introduce a
common, Gaussian constrained, scaling factor for the widths of all
three Gaussians, in order to account for the uncertainty in the width
of the resolution function.

The signal PDF for the $\Lambda_c(2625)^+$ is the nonrelativistic
Breit-Wigner function of Eq.~\ref{eq:BW} convolved with a three
Gaussian resolution function determined from simulation, which has an
average width of about 2.4\,\mevcc. Again, all three Gaussians have
mean zero and a common, Gaussian constrained, scaling factor for their
widths is introduced.

\begin{figure}
  \centering
  \includegraphics[width=8cm]{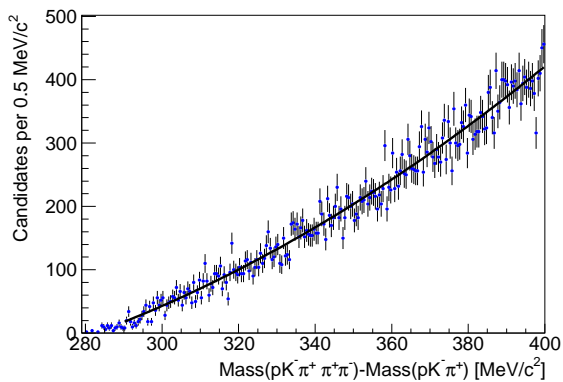}
  \caption {Fit to the $M(pK^-\pi^+\,\pi^+\pi^-)-M(pK^-\pi^+)$
    distribution of the candidates from $\Lambda_c^+$ mass sidebands.}
  \label{fig:LcS_LcSidebandFit}
\end{figure}

\begin{figure*}
  \centerline{\includegraphics[width=8cm]{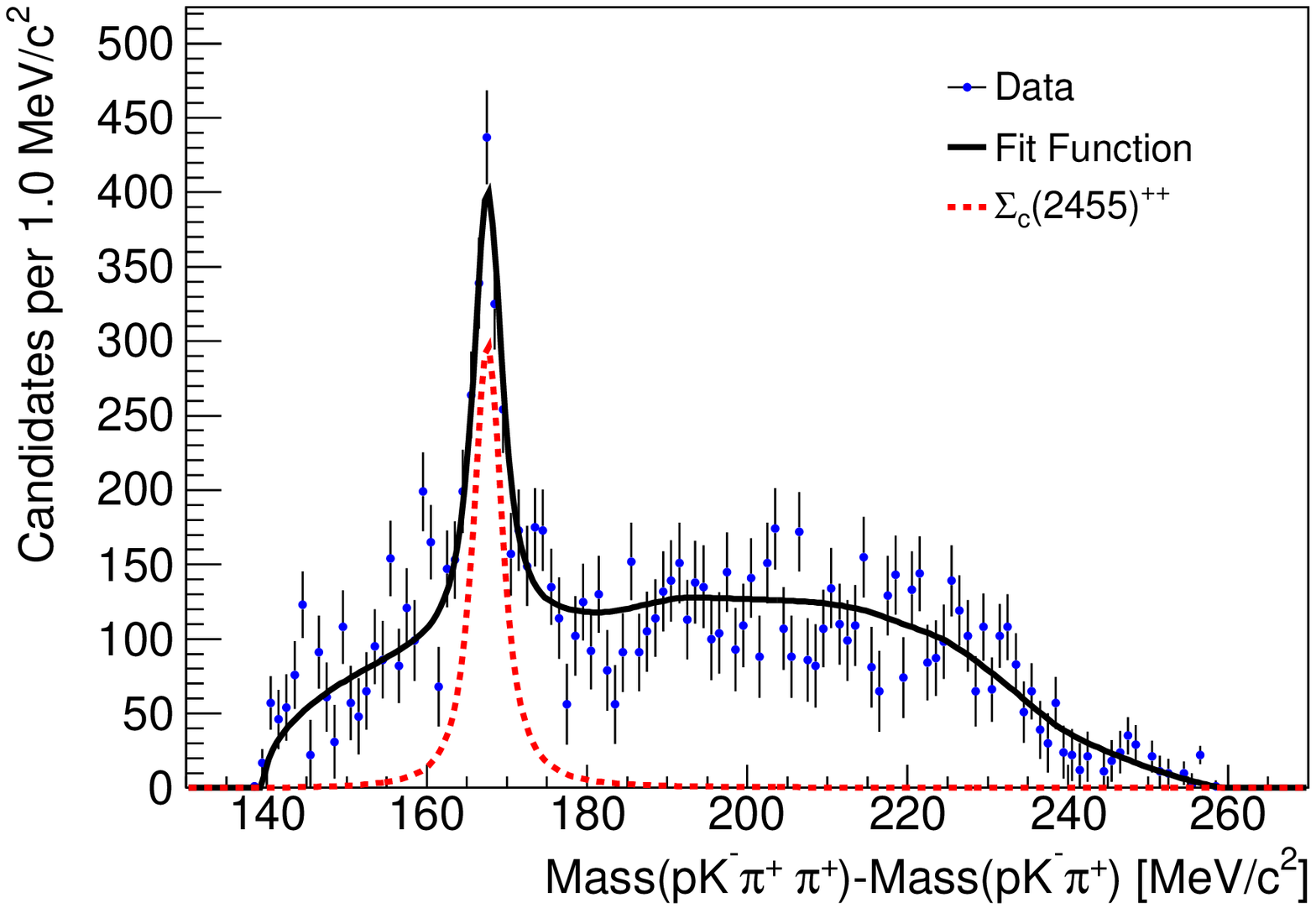}
    \includegraphics[width=8cm]{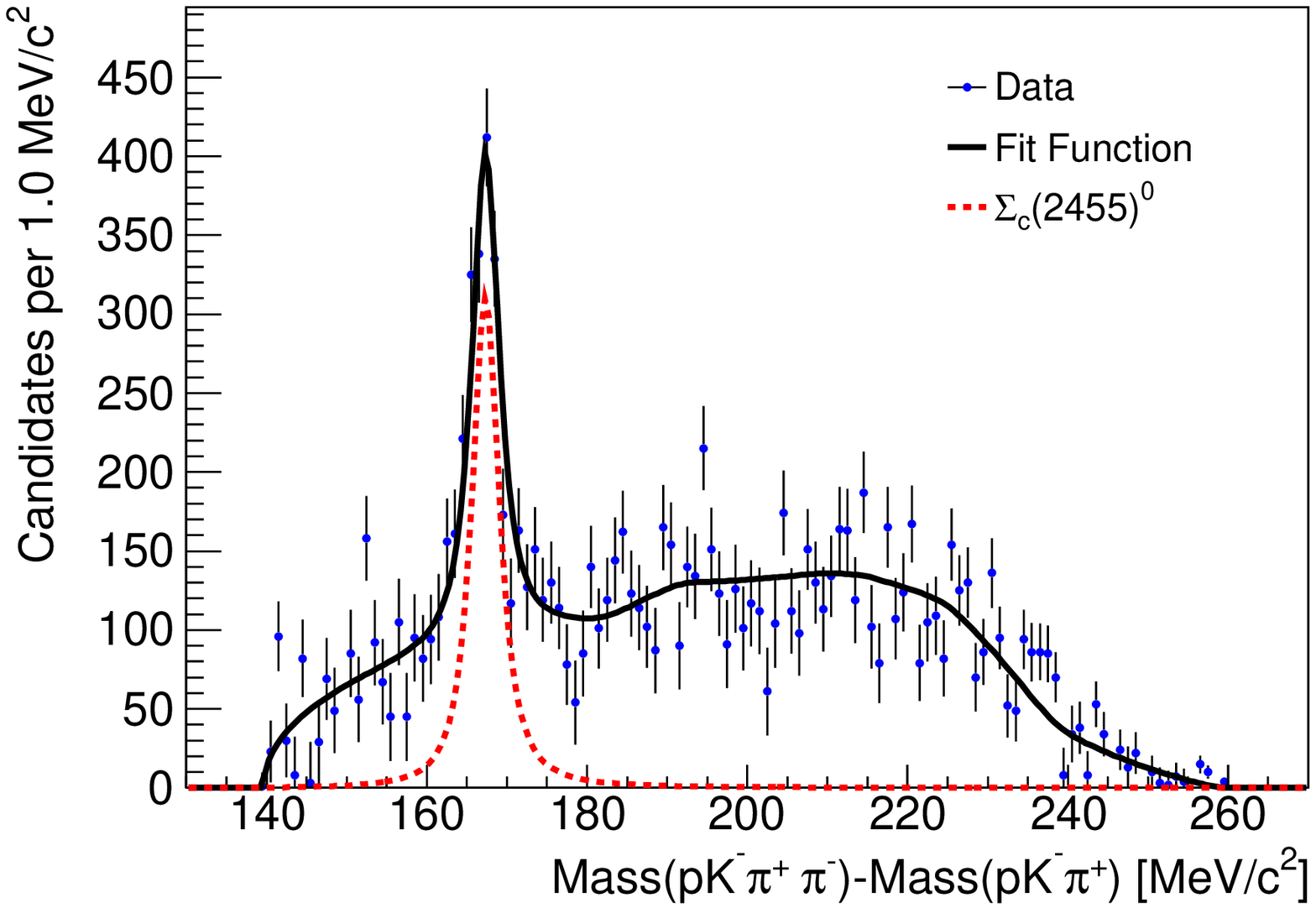}}
  \caption{(color online) Distributions of
    $M(\Sigma_c^{++})-M(pK^-\pi^+)$ (left) and
    $M(\Sigma_c^0)-M(pK^-\pi^+)$ (right) for candidates with
    $M(\Lambda_c^{*+})-M(pK^-\pi^+)>355$\,\mevcc together with the
    fits.}
  \label{fig:ScFit_LcS_sideband}
\end{figure*}

The background consists of three different sources, which include
combinatorial background without real $\Lambda_c^+$, real
$\Lambda_c^+$ combined with two random pions, and real
$\Sigma_c^{++,0}$ combined with a random pion. The combinatorial
background without real $\Lambda_c^+$ is parametrized by a
second-order polynomial whose parameters are determined in a fit to
the $\Delta M$ distribution of candidates from the $\Lambda_c^+$ mass
sidebands, $2261.46 < M(pK^-\pi^+) < 2266.46\,\mevcc$ and $2306.46 <
M(pK^-\pi^+) < 2311.46\,\mevcc$. This distribution is shown in
Fig.~\ref{fig:LcS_LcSidebandFit} together with the fit. In the final
fit, we keep the parameters for this background floating within a
Gaussian constraint of the form of Eq.~\ref{eq:GCMatrix} to the values
found in the fit to the candidates from the $\Lambda_c^+$ mass
sidebands. The second source, consisting of real $\Lambda_c^+$
combined with two random pions, is parametrized by a second-order
polynomial with all parameters allowed to float in the fit. The final
source of background are real $\Sigma_c$ combined with a random pion.
For this source, the main issue is to have the proper shape close to
the threshold. Small imperfections at higher $\Delta M$ can be
ignored, as the second background source has enough flexibility to
absorb it. The PDF of this $\Sigma_c$ background is based on a
constant function defined from the threshold to the end of the fit
range. In order to take into account the natural widths as well as
resolution effects, we use the weighted sum of ten such functions for
both $\Sigma_c(2455)^{++}$ and $\Sigma_c(2455)^0$. Their thresholds
and weights are chosen according to the shapes derived in the
$\Sigma_c$ fits shown in Fig.~\ref{fig:Sc_Fit}. The size of this
contribution is constrained to the $\Sigma_c(2455)$ yield obtained
from the fits to the $M(\Sigma_c)-M(pK^-\pi^+)$ distributions for
candidates with $M(\Lambda_c^{*+})-M(pK^-\pi^+)>355$\,\mevcc. These
two distributions together with the fits are shown in
Fig.~\ref{fig:ScFit_LcS_sideband}.

\begin{figure}
  \centering\includegraphics[width=8cm]{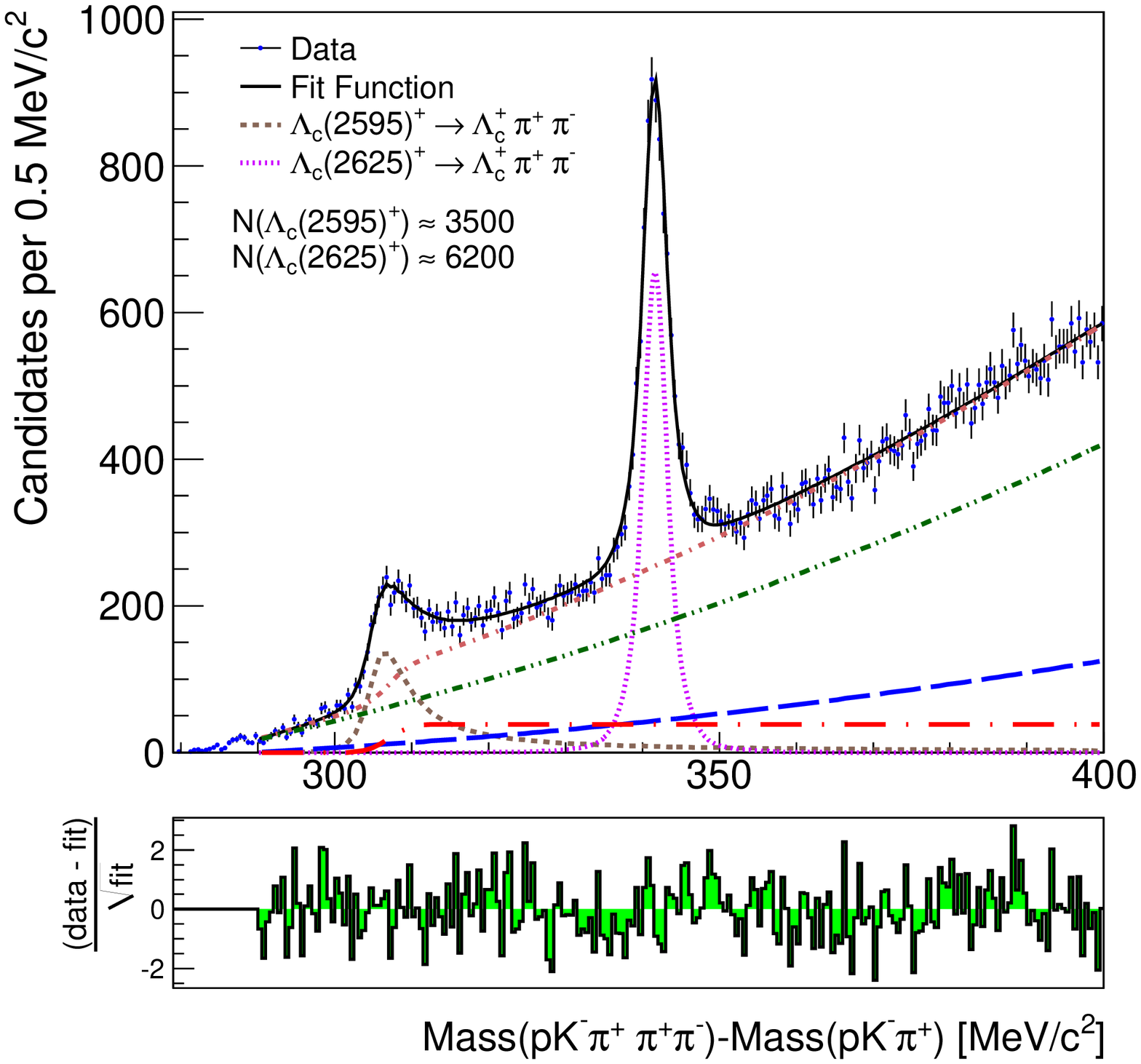}
  \caption{(color online) The $M(pK^-\pi^+\,\pi^+\pi^-)-M(pK^-\pi^+)$
    distribution obtained from data (points with error bars) together
    with the fit (black solid line). The brown dashed and purple
    dotted lines correspond to the two signal contributions, the green
    dash-double-dotted line represents the combinatorial background
    without real $\Lambda_c^+$, the blue long-dashed line shows real
    $\Lambda_c^+$ combined with two random pions and the red
    long-dash-dotted line represents real $\Sigma_c$ combined with a
    random pion. The red dash-dotted line corresponds to the sum of
    all three background contributions.}
  \label{fig:LcS_Fit}
\end{figure}

\begin{figure}
  \centering\includegraphics[width=8cm]{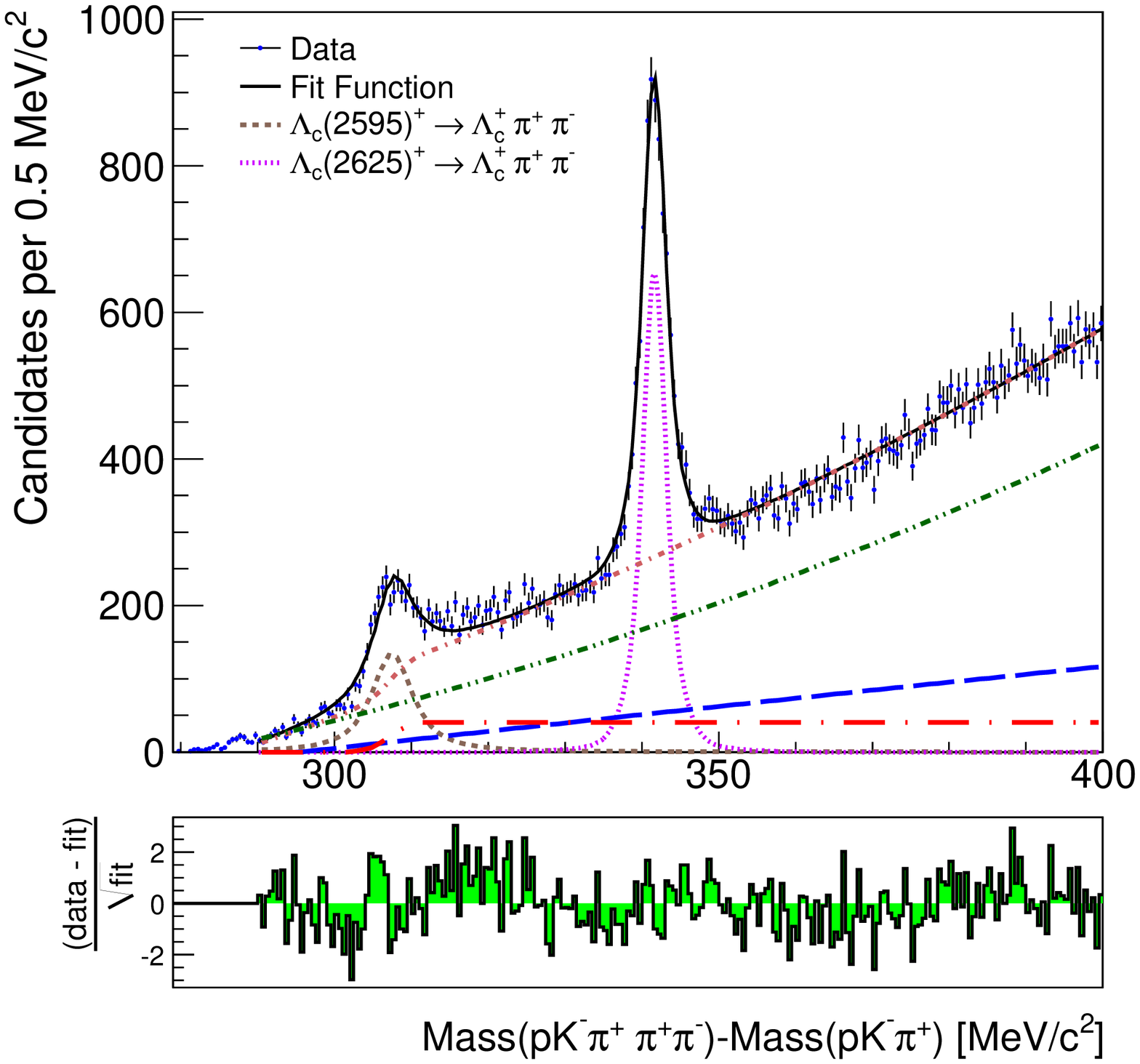}
  \caption{(color online) The $M(pK^-\pi^+\,\pi^+\pi^-)-M(pK^-\pi^+)$
    distribution obtained from data (points with error bars) together
    with the fit (black solid line), where a Breit-Wigner function
    with a mass-independent decay width is used to model the
    $\Lambda_c(2595)^+$ line shape. Explanations of the various
    background contributions can be found in the caption of
    Fig.~\ref{fig:LcS_Fit}.}
  \label{fig:LcS_FitUsualBW}
\end{figure}

The full fit to the $\Delta M$ distribution, containing all signal and
background components, can be found in Fig.~\ref{fig:LcS_Fit}. The
$\chi^2$ value of the fit is 227 (206 degrees of freedom). Compared to
that, the $\chi^2$ value of a fit with a mass-independent
$\Lambda_c(2595)^+$ decay width, shown in
Fig.~\ref{fig:LcS_FitUsualBW}, increases to 286 (206 degrees of
freedom).

\section{Systematic uncertainties}
\label{sec:systematics}

We investigate several systematic effects that can affect the
measurements. Generally, they can be categorized as imperfect modeling
by the simulation, imperfect knowledge of the momentum scale of the
detector, ambiguities in the fit model, and uncertainties on the
external inputs to the fit. In this Section we discuss how they can
affect our results and the way we assess them. A summary of the
assigned uncertainties can be found in
Tables~\ref{tab:Sc++_Syst}--\ref{tab:LcS_Syst}. To obtain the total
systematic uncertainties, we add up the contributions from all sources
in quadrature.

\begin{table*}
  \centering
  \caption{Systematic uncertainties on the
    measurements of the mass differences and decay widths of
    the $\Sigma_c^{++}$ resonances. The corresponding statistical
    uncertainties are listed for comparison.}
  \begin{tabular}{lcccc}
    \hline
    \hline
    Source & $\Delta M(\Sigma_c(2455)^{++})$ & $\Gamma(\Sigma_c(2455)^{++})$ & $\Delta M(\Sigma_c(2520)^{++})$ & $\Gamma(\Sigma_c(2520)^{++})$\\
    & [\mevcc] & [\mevcc] & [\mevcc] & [\mevcc]\\
    \hline
    Resolution model & $\cdot\cdot\cdot$ & 0.40 & $\cdot\cdot\cdot$ & 0.69\\
    Momentum scale & 0.12 & 0.20 & 0.12 & 0.20\\
    Fit model & 0.02 & $\cdot\cdot\cdot$ & 0.11 & 1.16\\
    External inputs & $\cdot\cdot\cdot$ & $\cdot\cdot\cdot$ & $\cdot\cdot\cdot$ & $\cdot\cdot\cdot$\\
    \hline
    Sum & 0.12 & 0.45 & 0.16 & 1.36\\
    \hline
    Statistical & 0.04 & 0.13 & 0.56 & 2.12\\
    \hline
    \hline
  \end{tabular}
  \label{tab:Sc++_Syst}
\end{table*}

\begin{table*}
  \centering
  \caption{Systematic uncertainties on the measurements
    of the mass differences and decay widths of the $\Sigma_c^0$
    resonances. The corresponding statistical uncertainties are
    listed for comparison.}
  \begin{tabular}{lcccc}
    \hline\hline
    Source & $\Delta M(\Sigma_c(2455)^0)$ & $\Gamma(\Sigma_c(2455)^0)$ & $\Delta M(\Sigma_c(2520)^0)$ & $\Gamma(\Sigma_c(2520)^0)$\\
    & [\mevcc] & [\mevcc] & [\mevcc] & [\mevcc]\\
    \hline
    Resolution model & $\cdot\cdot\cdot$ & 0.45 & $\cdot\cdot\cdot$ & 0.70\\
    Momentum scale & 0.12 & 0.20 & 0.12 & 0.20\\
    Fit model & 0.02 & $\cdot\cdot\cdot$ & 0.11 & 1.16\\
    External inputs & $\cdot\cdot\cdot$ & $\cdot\cdot\cdot$ & $\cdot\cdot\cdot$ & $\cdot\cdot\cdot$\\
    \hline
    Sum & 0.12 & 0.49 & 0.16 & 1.37\\
    \hline
    Statistical & 0.03 & 0.11 & 0.43 & 1.82\\
    \hline
    \hline
  \end{tabular}
  \label{tab:Sc0_Syst}
\end{table*}

\begin{table*}
  \centering
  \caption{Systematic uncertainties on the measurements
    of the mass differences of the $\Lambda_c^{*+}$ resonances and the
    pion coupling constant $h_2^2$ ($\Gamma(\Lambda_c(2595)^+)$).
    The corresponding statistical uncertainties are listed for comparison.}
  \begin{tabular}{lcccc}
    \hline
    \hline
    Source & $\Delta M(\Lambda_c(2595)^+)$ & $h_2^2$ & $\Gamma(\Lambda_c(2595)^+)$ & $\Delta M(\Lambda_c(2625)^+)$\\
    & [\mevcc] &   & [\mevcc] & [\mevcc]\\
    \hline
    Resolution model & 0.06 & 0.03 & 0.22 & $\cdot\cdot\cdot$\\
    Momentum scale & 0.12 & 0.03 & 0.20 & 0.12\\
    Fit model & $\cdot\cdot\cdot$ & $\cdot\cdot\cdot$ & $\cdot\cdot\cdot$ & $\cdot\cdot\cdot$\\
    External inputs & 0.15 & 0.06 & 0.36 & $\cdot\cdot\cdot$\\
    \hline
    Sum & 0.20 & 0.07 & 0.47 & 0.12\\
    \hline
    Statistical & 0.14 & 0.04 & 0.30 & 0.04\\
    \hline
    \hline
  \end{tabular}
  \label{tab:LcS_Syst}
\end{table*}

\subsection{Mass resolution model}

\begin{table}
  \centering
  \caption{Mass resolution scaling factors $s$ floating within Gaussian constraints in the fits.}
  \begin{tabular}{cc}
    \hline
    \hline
    Hadron & $s$ \\
    \hline
    $\Sigma_c(2455)^{++}$ & $0.93 \pm 0.17$\\
    $\Sigma_c(2455)^{0}$  & $1.07 \pm 0.13$\\
    $\Sigma_c(2520)^{++}$ & $1.02 \pm 0.20$\\
    $\Sigma_c(2520)^{0}$  & $1.00 \pm 0.20$\\
    $\Lambda_c(2595)^{+}$ & $0.95 \pm 0.15$\\
    \hline
    \hline
  \end{tabular}
  \label{tab:scaling_factors}
\end{table}

To properly describe the signal shapes, we need to understand the
intrinsic mass resolution of the detector. Since we estimate this
using simulated events, it is necessary to verify that the resolution
obtained from simulation agrees with that in real data. We use
$D^*(2010)^+ \to D^0\,\pi^+$ with $D^0 \to K^-\,\pi^+$ decays and
$\psi(2S) \to \Jpsi\,\pi^+\,\pi^-$ with $\Jpsi \to \mu^+\,\mu^-$
decays for this purpose. We compare the resolution in data and
simulated events as a function of the $p_T$ of the pions added to
$D^0$ or $\Jpsi$ as well as the instantaneous luminosity. We also
compare the overall resolution scale between data and simulated events
and find that all discrepancies are less than 20\%, which we assign as
uncertainty on our knowledge of the resolution function. The
contribution from this uncertainty is already included in the
uncertainties on the resonance parameters determined by the default
fit with Gaussian constraint on the resolution scaling factor $s$, the
resulting values for which are listed in
Table~\ref{tab:scaling_factors}. These values are consistent with 1,
indicating that the resolution is well understood within the assigned
uncertainty. To disentangle it from the statistical component, we
repeat the fits on data without multiplying the widths of the
resolution function by the scaling factor $s$ from Eq.~\ref{eq:GC}.
The systematic uncertainty due to the imperfect modeling of the
resolution function is then obtained by the difference in quadrature
of the uncertainty of the fit with and without the Gaussian
constraint. This uncertainty in the resolution has a large impact on
the natural widths, but a negligible effect on the mass differences.

\subsection{Momentum scale}

The accuracy of the momentum scale depends on the precision with which
the magnetic field and the amount of material in the detector are
known. Both effects are originally calibrated using $\Jpsi \to
\mu^+\,\mu^-$ decays~\cite{Acosta:2005}. A limitation of this
calibration is that it uses muons that are required by the detector
acceptance to have $p_T>1.5$\,\gevc, while pions from $\Sigma_c$ or
$\Lambda_c^{*+}$ decays typically have much lower $p_T$. The estimate
of the uncertainty on the mass differences comes from our previous
work on the $X(3872)$ hadron~\cite{Aaltonen:2009vj}. There, $\psi(2S)
\to \Jpsi\,\pi^+\,\pi^-$ decays are used to study the momentum scale
uncertainties by comparing the measured $\psi(2S)$ mass with the world
average value~\cite{PDG}. In addition, we study the $\psi(2S)$ mass
dependence on the kinematic properties of the pions, which constrains
the sizes of possible effects. Furthermore, we verify the momentum
scale by using $D^*(2010)^+ \to D^0\,\pi^+$ decays, where the
resulting deviation from the world average is far below the
uncertainty derived from $\psi(2S)$. Based on
Ref.~\cite{Aaltonen:2009vj}, we assign a 0.12\,\mevcc uncertainty on
the mass differences of all states under study due to the imperfect
knowledge of the momentum scale. The corresponding effect on the
natural widths was studied in our previous measurements of the masses
and widths of the excited charmed meson states $D_1^0$ and
$D_2^{*0}$~\cite{Abulencia:2005ry}, and we assign the 0.2\,\mevcc
found there as the uncertainty on the natural widths due to this
source. To translate this uncertainty to the coupling constant $h_2$,
we assign it to the sum $\Gamma(\Lambda_c^+\pi^+\pi^-) +
\Gamma(\Lambda_c^+\pi^0\pi^0)$ (see Eqs.~\ref{eq:Lc(2595)_Gamma1} and
\ref{eq:Lc(2595)_Gamma2}), which is a function of $h_2$, and perform
Gaussian error propagation.

\subsection{Fit model}

In terms of our fit model and procedure we check two effects, the
internal consistency of the fit and the shape of the signal PDFs. We
do not perform an explicit check of the background parametrizations as
those are described by polynomials and any analytic function can be
approximated by a polynomial of sufficient complexity. Since the fit
quality does not indicate significant discrepancies between data and
the model, we conclude that the degree of the polynomial functions
used is sufficient. Some backgrounds are determined from independent
sources, but as the appropriate parameters are Gaussian constrained in
the fit, the uncertainty originating from the sample size of the
external sources, like $\Lambda_c^+$ mass sidebands, is already
included in the statistical uncertainties of the results.

To check the internal consistency of the fit procedure, we generate a
large ensemble of statistical trials using PDFs of our fit model with
parameters obtained from the fit to data. Estimates of all physics
parameters except the mass differences and natural widths of the
$\Sigma_c(2520)$ resonances are found to be unbiased. The
$\Sigma_c(2520)$ mass differences have small biases towards higher
values and the $\Sigma_c(2520)$ natural widths are biased towards
lower values. These biases on the $\Sigma_c(2520)$ resonance
parameters result from the fairly low signal to background ratio and
the flexibility in the background PDF, which tends to absorb the tails
of the relatively broad signal structure. We repeat the study with a
true value for the $\Sigma_c(2520)$ natural width below
($\Gamma=7.5$\,\mevcc) and above ($\Gamma=20$\,\mevcc) the measured
value and find that the biases have a small dependence on the true
value. The biases are largest for a true value of the natural width of
20\,\mevcc and we consequently assign these biases as systematic
uncertainties on the mass differences and natural widths of the
$\Sigma_c(2520)$ states.

Concerning the uncertainty on the signal shape, we check whether our
signal parametrization using nonrelativistic Breit-Wigner functions
provides a proper description. We refit the $\Sigma_c$ and
$\Lambda_c(2625)^+$ data using a $P$-wave relativistic Breit-Wigner
function of the form
\begin{equation}
  \frac{\mathrm{d}N}{\mathrm{d}m} \propto \frac{m \cdot
    \Gamma(m)}{(m_0^2 - m^2)^2 + m_0^2 \cdot \Gamma^2(m)}
  \label{eq:RelBW}
\end{equation}
with
\begin{equation}
  \Gamma(m) = \Gamma_0 \left(\frac{q}{q_0}\right)^3
  \left(\frac{m_0}{m}\right) \left(\frac{1 + q_0^2 R^2}{1 +
      q^2 R^2}\right),
  \label{eq:Pwave}
\end{equation}
where $m = \Delta M + m_{\Lambda_c^+}$, $R$ is the Blatt-Weisskopf
radius set to
3\,$(\mathrm{GeV}/c)^{-1}$~\cite{Aston:1987ir,Godfrey:1985xj}, $m_0$
and $\Gamma_0$ are the nominal mass and width, and $q(q_0)$ is the
momentum of the daughters in the $\Sigma_c$ or $\Lambda_c(2625)^+$
rest frame calculated from the nominal mass. For the
$\Lambda_c(2595)^+$ we replace the nonrelativistic Breit-Wigner
function of Eq.~\ref{eq:Lc(2595)_BW} by a relativistic one and use the
variable width defined in Eqs.~\ref{eq:Lc(2595)_Gamma1} and
\ref{eq:Lc(2595)_Gamma2}. For the $\Sigma_c(2455)$ we observe a
difference of 0.02\,\mevcc in the mass difference, which we assign as
a systematic uncertainty. In the cases of $\Sigma_c(2520)$ and
$\Lambda_c^{*+}$ resonances we do not observe any shift and conclude
that the effect is negligible.

\subsection{External inputs}

Finally, the line shape of the $\Lambda_c(2595)^+$ depends on the
input values of the $\Sigma_c(2455)$ masses and widths and the pion
decay constant $f_\pi$. We repeat the fit using values of those
parameters smaller or larger by 1 standard deviation and take the
stronger variation as systematic uncertainty. The effect of the
uncertainty on the world average $\Sigma_c(2455)$ masses and widths
used as input is dominant compared to the effect of the uncertainty on
$f_\pi$.

\section{Results and conclusions}
\label{sec:results}

\begin{table}
  \centering
  \caption{Measured resonance parameters, where the first uncertainty is statistical and the second is systematic.}
  \begin{tabular}{ccc}
    \hline
    \hline
    Hadron & $\Delta M$ [\mevcc] & $\Gamma$ [\mevcc]\\
    \hline
    $\Sigma_c(2455)^{++}$ & $167.44 \pm 0.04 \pm 0.12$ & $2.34  \pm 0.13 \pm 0.45$\\
    $\Sigma_c(2455)^{0}$  & $167.28 \pm 0.03 \pm 0.12$ & $1.65  \pm 0.11 \pm 0.49$\\
    $\Sigma_c(2520)^{++}$ & $230.73 \pm 0.56 \pm 0.16$ & $15.03 \pm 2.12 \pm 1.36$\\
    $\Sigma_c(2520)^{0}$  & $232.88 \pm 0.43 \pm 0.16$ & $12.51 \pm 1.82 \pm 1.37$\\
    $\Lambda_c(2595)^{+}$ & $305.79 \pm 0.14 \pm 0.20$ & $h_2^2 = 0.36 \pm 0.04 \pm 0.07$\\
    $\Lambda_c(2625)^{+}$ & $341.65 \pm 0.04 \pm 0.12$ & \\
    \hline
    \hline
  \end{tabular}
  \label{tab:FitResults}
\end{table}

We perform fits to the $M(pK^-\pi^+\,\pi^+)-M(pK^-\pi^+)$,
$M(pK^-\pi^+\,\pi^-)-M(pK^-\pi^+)$, and
$M(pK^-\pi^+\,\pi^+\pi^-)-M(pK^-\pi^+)$ mass difference distributions
to obtain the desired resonance properties. The data distributions and
fits are shown in Figs.~\ref{fig:Sc_Fit} and \ref{fig:LcS_Fit}. We
select about 13800 $\Sigma_c(2455)^{++}$, 15900 $\Sigma_c(2455)^{0}$,
8800 $\Sigma_c(2520)^{++}$, 9000 $\Sigma_c(2520)^{0}$, 3500
$\Lambda_c(2595)^+$, and 6200 $\Lambda_c(2625)^+$ signal events. The
resonance parameters obtained can be found in
Table~\ref{tab:FitResults}. For the width of the $\Lambda_c(2625)^+$
we observe a value consistent with zero and therefore calculate an
upper limit using a Bayesian approach with a uniform prior restricted
to positive values. At the 90\% credibility level we obtain
$\Gamma(\Lambda_c(2625)^+)<0.97$\,\mevcc.  For easier comparison to
previous results~\cite{Albrecht:1997qa,Edwards:1994ar}, $h_2^2$
corresponds to a $\Lambda_c(2595)^+$ decay width of
$\Gamma(\Lambda_c(2595)^+)=2.59 \pm 0.30 \pm 0.47$\,\mevcc, calculated
at $\Delta M(\Lambda_c(2595)^+)$. Our precise measurement of the
coupling constant $h_2$ can, for instance, be used to predict the
width of the $\Xi_c(2645)$, as discussed in
Ref.~\cite{Chiladze:1997ev}.

\begin{figure*}
  \centerline{\includegraphics[width=6.0cm]{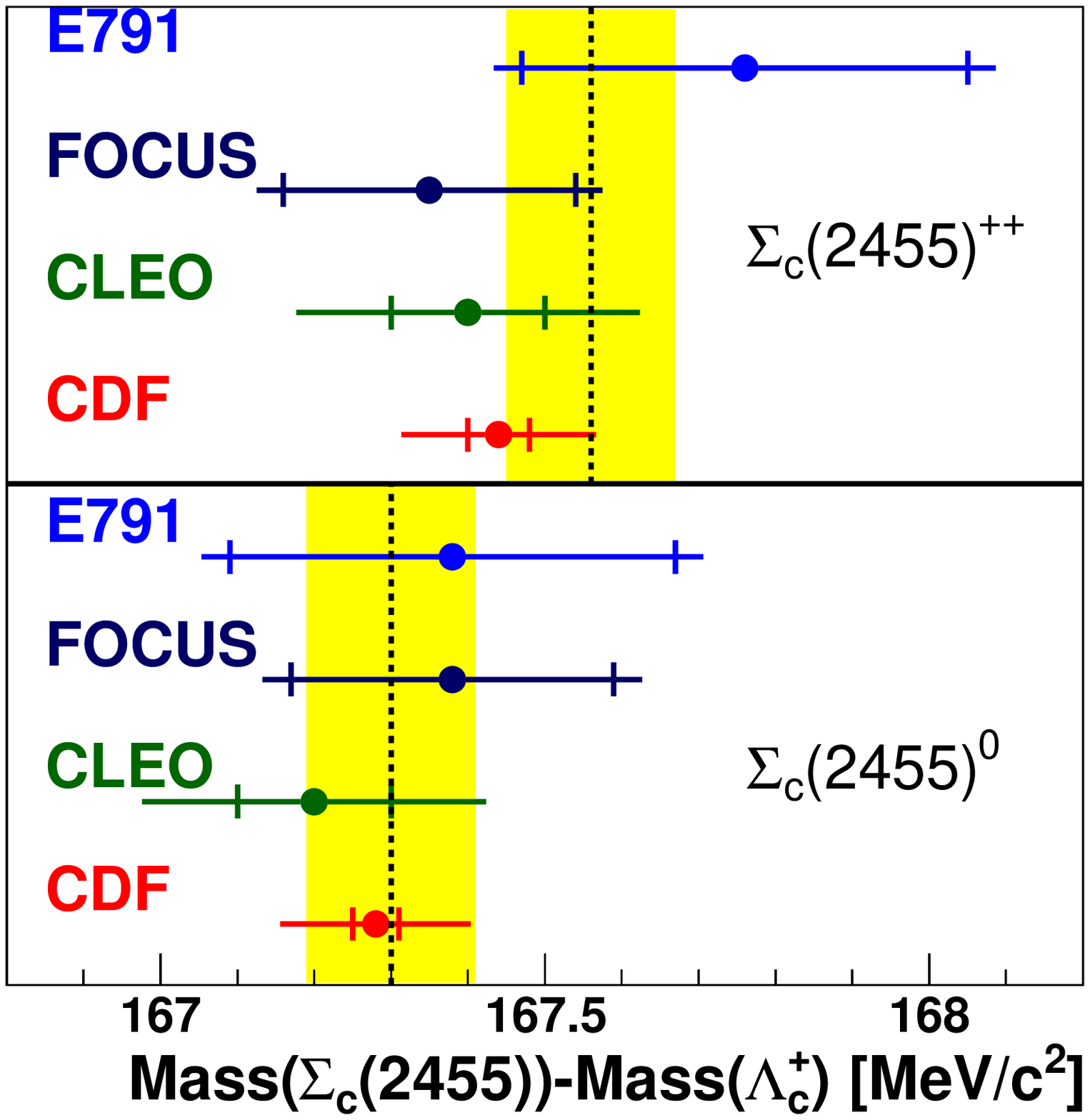}
    \includegraphics[width=6.0cm]{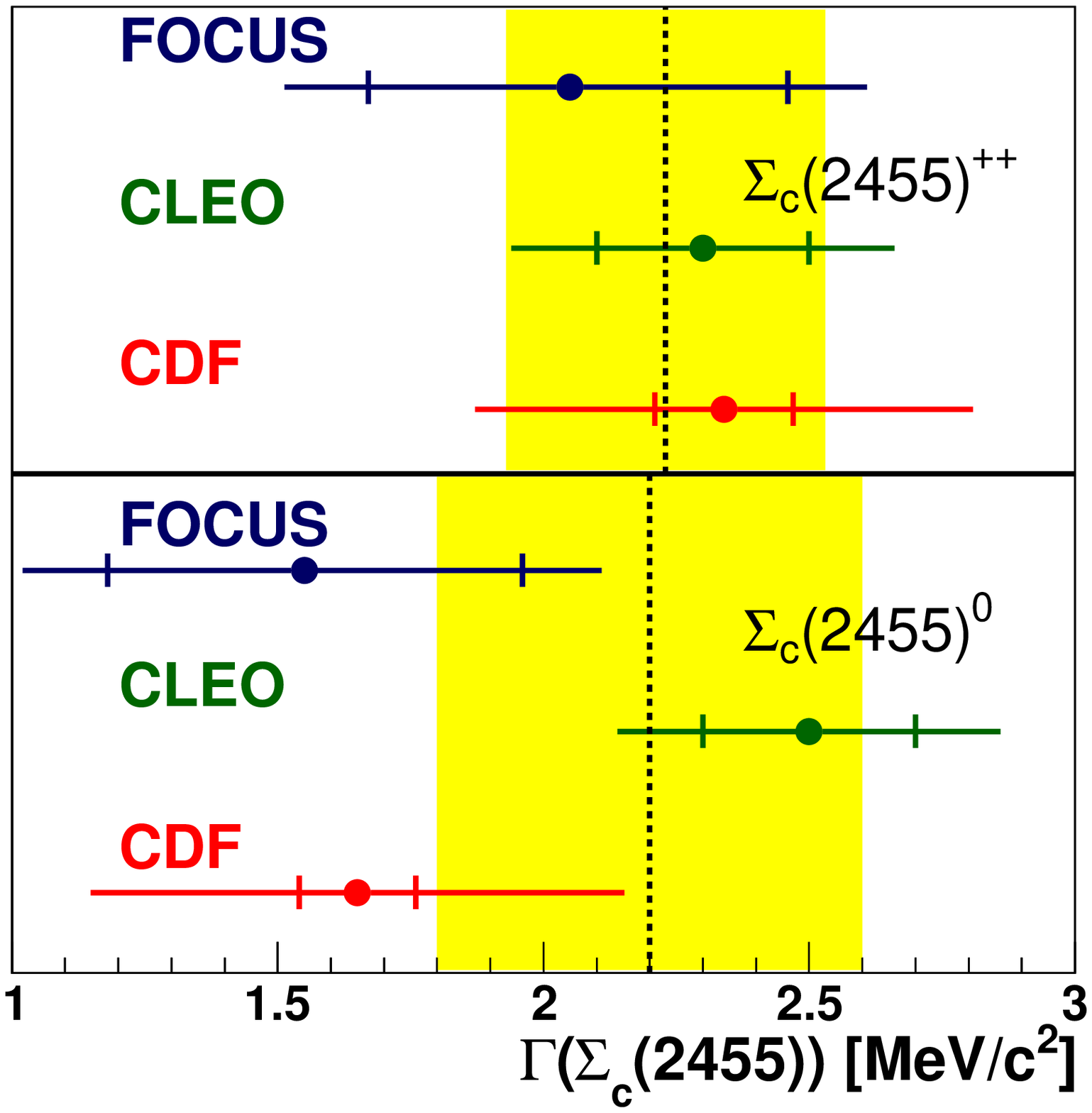}}
  \caption{Comparison of our results for the $\Sigma_c(2455)$ mass
    differences and decay widths with previous measurements by
    Fermilab E791~\cite{Aitala:1996}, FOCUS~\cite{Link:2001ee}, and
    CLEO~\cite{Artuso:2001us}.  The error bars represent the
    statistical (vertical marks) as well as the combined statistical
    and systematic uncertainties. The vertical dashed line together
    with the surrounding box symbolizes the world average value and
    its uncertainty taken from Ref.~\cite{PDG}. This average does not
    take into account the measurement at hand.}
  \label{fig:Sc2455Comp}
\end{figure*}

\begin{figure*}
  \centerline{\includegraphics[width=6.0cm]{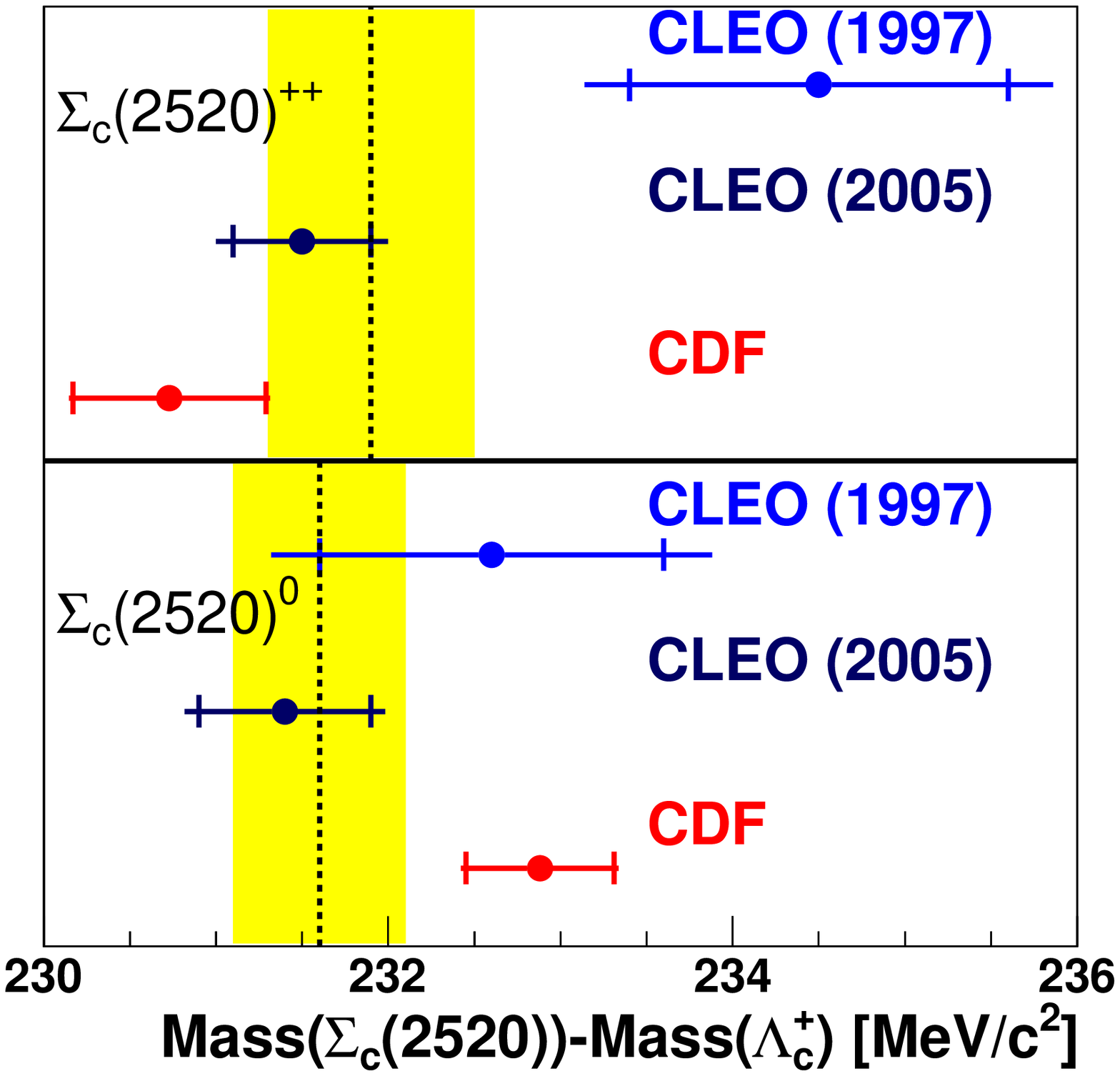}
    \includegraphics[width=6.0cm]{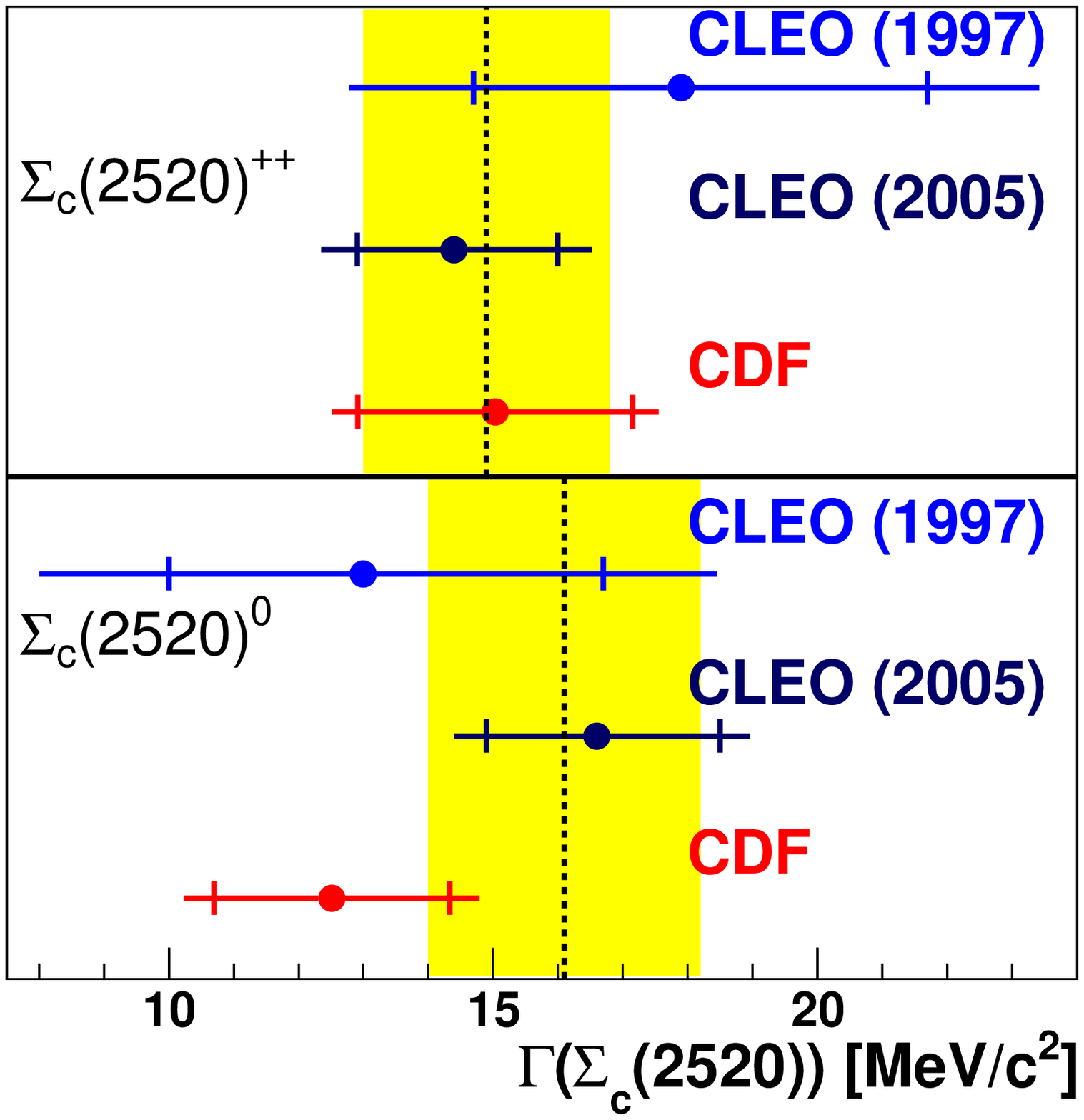}}
  \caption{Comparison of our results for the $\Sigma_c(2520)$ mass
    differences and decay widths with previous measurements by
    CLEO~\cite{Brandenburg:1996jc,Athar:2004ni}. Further explanations
    can be found in the caption of Fig.~\ref{fig:Sc2455Comp}.}
  \label{fig:Sc2520Comp}
\end{figure*}

\begin{figure*}
  \centerline{\includegraphics[width=6.0cm]{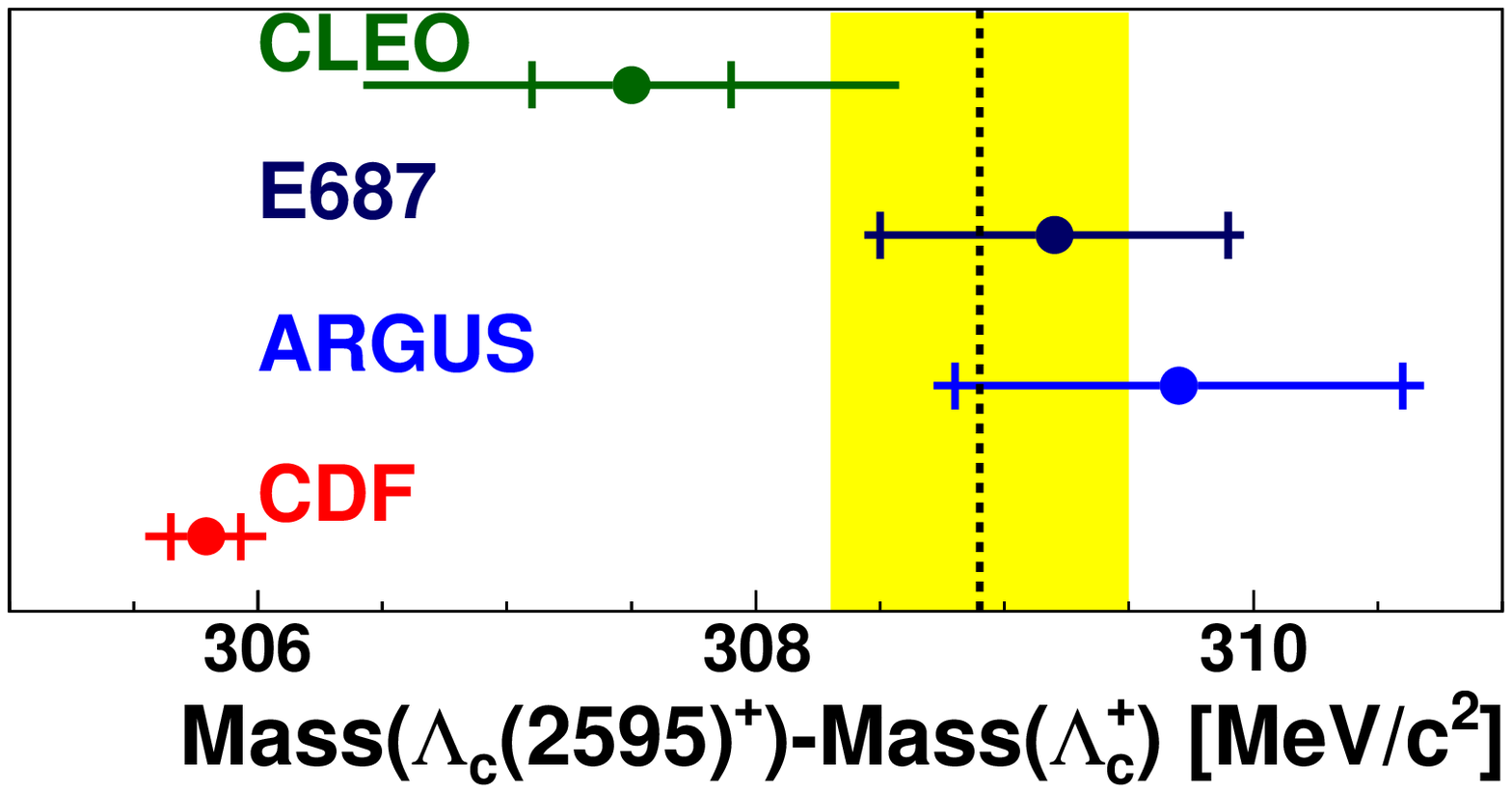}
    \includegraphics[width=6.0cm]{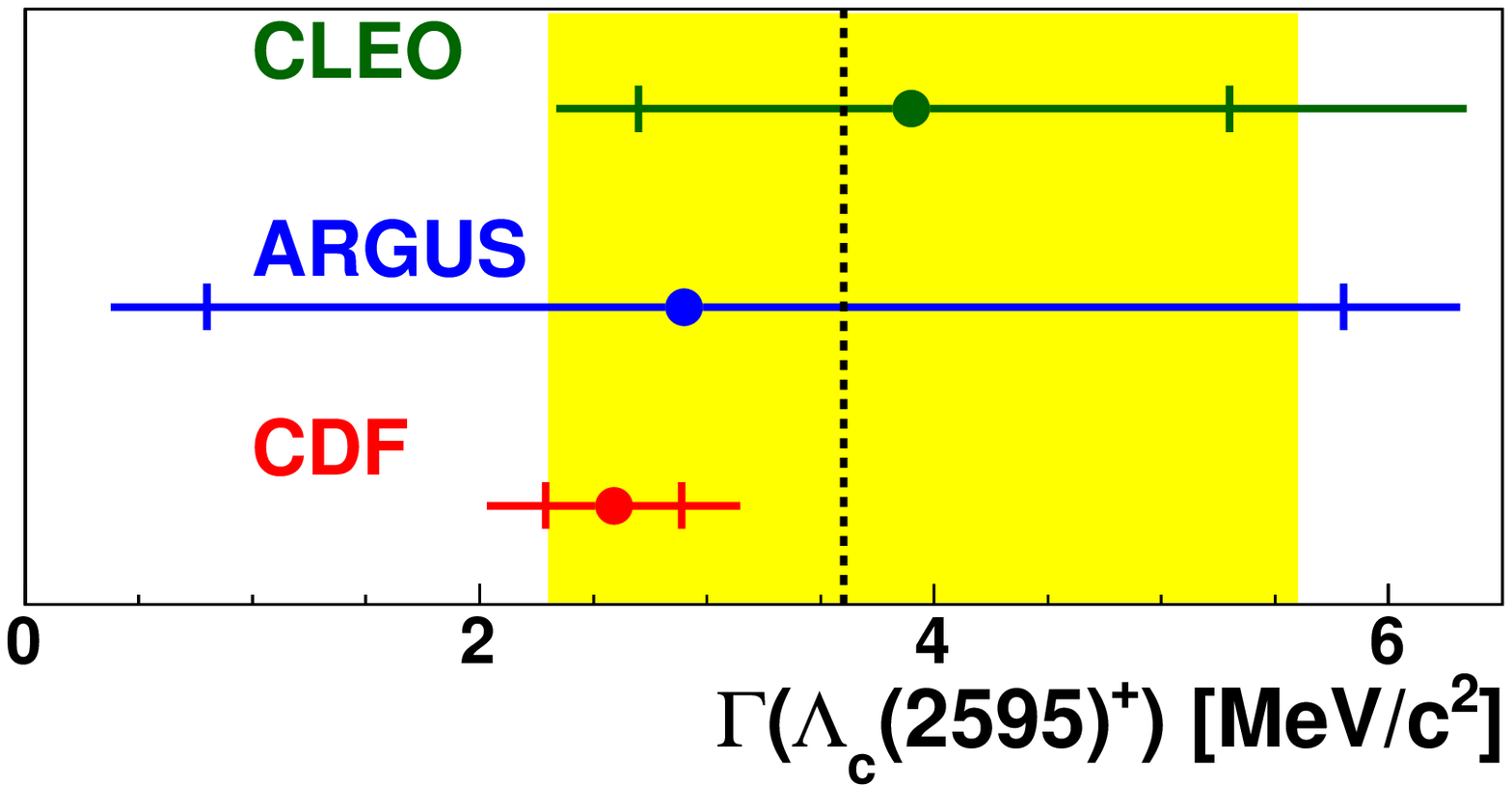}
    \includegraphics[width=6.0cm]{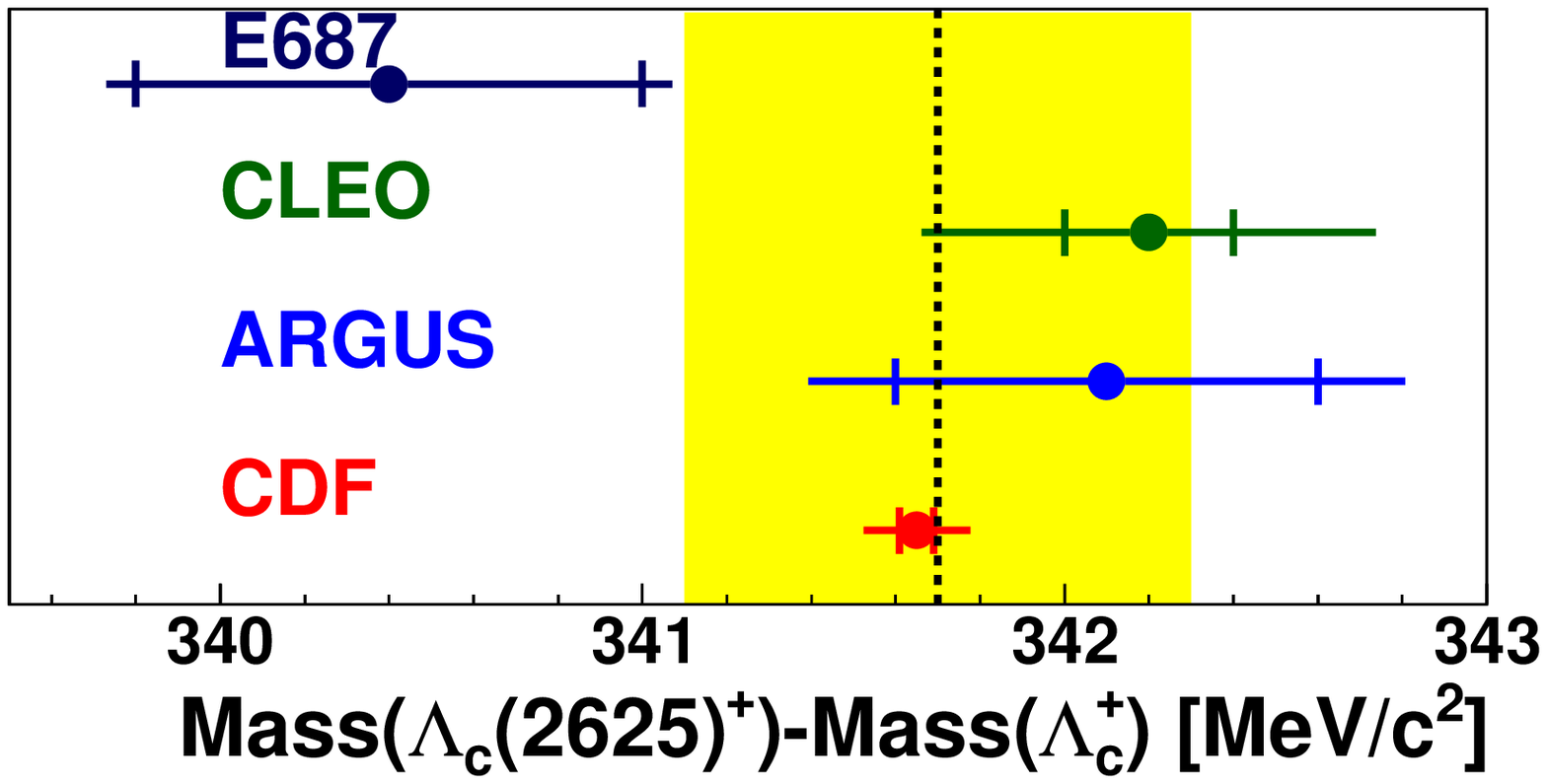}}
  \caption{Comparison of our results for the $\Lambda_c(2595)^+$ mass
    difference and decay width as well as the $\Lambda_c(2625)^+$ mass
    difference with previous measurements by CLEO
    ~\cite{Edwards:1994ar}, Fermilab
    E687~\cite{Frabetti:1994,Frabetti:1995sb}, and
    ARGUS~\cite{Albrecht:1997qa}. Further explanations can be found in
    the caption of Fig.~\ref{fig:Sc2455Comp}.}
  \label{fig:LcSComp}
\end{figure*}

In Figs.~\ref{fig:Sc2455Comp}--\ref{fig:LcSComp}, our results are
compared to previous measurements by other experiments. Except for
$\Delta M(\Lambda_c(2595)^+)$, all our measurements agree with the
previous world average values. For $\Delta M(\Lambda_c(2595)^+)$ we
show that a mass-independent natural width does not describe the data
(see Fig.~\ref{fig:LcS_FitUsualBW}) and observe a value which is
3.1\,\mevcc smaller than the existing world average. This difference
is the same size as estimated in Ref.~\cite{Blechman:2003mq}. Since
this data sample is 25 times larger than the ones studied so far, our
results on the properties of $\Lambda_c^{*+}$ states provide a
significant improvement in precision compared to previous
measurements. The precision for the $\Sigma_c$ states is comparable to
the precision of the world averages. Concerning the inconsistency of
the two CLEO measurements~\cite{Brandenburg:1996jc,Athar:2004ni} of
the $\Sigma_c(2520)^{++}$ mass, our data favor a smaller value.

In conclusion, we exploit the world largest samples of excited charmed
baryons to measure the resonance parameters of six states, namely
$\Sigma_c(2455)^{++}$, $\Sigma_c(2455)^{0}$, $\Sigma_c(2520)^{++}$,
$\Sigma_c(2520)^{0}$, $\Lambda_c(2595)^+$, and $\Lambda_c(2625)^+$.
Table~\ref{tab:FinalResults} summarizes the results for their masses
and widths. These measurements provide a significant improvement in
the knowledge of the resonance parameters of the states and represent
the first analysis of charmed baryons at a hadron collider.

\begin{table}
  \centering
  \caption{Results for the masses and widths of the charmed baryons under study. The first uncertainty is the combined statistical and systematic experimental uncertainty. For the masses, the second uncertainty originates from the world average $\Lambda_c^+$ mass~\cite{PDG}.}
  \begin{tabular}{ccc}
    \hline
    \hline
    Hadron & $M$ [\mevcc] & $\Gamma$ [\mevcc]\\
    \hline
    $\Sigma_c(2455)^{++}$ & $2453.90 \pm 0.13 \pm 0.14$ & $2.34  \pm 0.47$\\
    $\Sigma_c(2455)^{0}$  & $2453.74 \pm 0.12 \pm 0.14$ & $1.65  \pm 0.50$\\
    $\Sigma_c(2520)^{++}$ & $2517.19 \pm 0.46 \pm 0.14$ & $15.03 \pm 2.52$\\
    $\Sigma_c(2520)^{0}$  & $2519.34 \pm 0.58 \pm 0.14$ & $12.51 \pm 2.28$\\
    $\Lambda_c(2595)^{+}$ & $2592.25 \pm 0.24 \pm 0.14$ & $h_2^2 = 0.36 \pm 0.08$\\
    $\Lambda_c(2625)^{+}$ & $2628.11 \pm 0.13 \pm 0.14$ & $<0.97\, \mathrm{at\,\, 90\%\,\, C.L.}$\\
    \hline
    \hline
  \end{tabular}
  \label{tab:FinalResults}
\end{table}

\begin{acknowledgments}
  We thank the Fermilab staff and the technical staffs of the
  participating institutions for their vital contributions. This work
  was supported by the U.S. Department of Energy and National Science
  Foundation; the Italian Istituto Nazionale di Fisica Nucleare; the
  Ministry of Education, Culture, Sports, Science and Technology of
  Japan; the Natural Sciences and Engineering Research Council of
  Canada; the National Science Council of the Republic of China; the
  Swiss National Science Foundation; the A.P. Sloan Foundation; the
  Bundesministerium f\"ur Bildung und Forschung, Germany; the Korean
  World Class University Program, the National Research Foundation of
  Korea; the Science and Technology Facilities Council and the Royal
  Society, UK; the Institut National de Physique Nucleaire et Physique
  des Particules/CNRS; the Russian Foundation for Basic Research; the
  Ministerio de Ciencia e Innovaci\'{o}n, and Programa
  Consolider-Ingenio 2010, Spain; the Slovak R\&D Agency; the Academy
  of Finland; and the Australian Research Council (ARC). We thank
  Andrew Blechman for providing feedback on the calculation of the
  $\Lambda_c(2595)^+$ line shape.
\end{acknowledgments}

\bibliography{cdfpubnote}

\end{document}